\documentclass[acmsmall, screen]{acmart}

\setcopyright{none}

\renewcommand\footnotetextcopyrightpermission[1]{}  
\AtBeginDocument{%
  }

\usepackage{booktabs}
\usepackage[table]{xcolor}
\usepackage{subcaption} 
\usepackage{xspace}
\usepackage{multirow}
\usepackage[varqu]{zi4}
\usepackage{pifont}

\newcommand{\sol}{{\textsc{NetGuise}}\xspace}
\newcommand{\GapOne}{\ding{202}\enspace}
\newcommand{\GapTwo}{\ding{203}\enspace}
\newcommand{\GapThree}{\ding{204}\enspace}

\newcommand{\msp}[2]{#1{\scriptsize$\pm$#2}}
\newcommand{\mspg}[3]{\textbf{#1{\scriptsize$\pm$#2}}{\tiny\textcolor{cyan!50!blue}{(#3)}}}

\begin{document}

\title{\textsc{NetGuise}: Adversarial Traffic Generation for Early-Stage Classifier Evasion}


\author{Binghui Wu}
\correspondingauthor
\affiliation{%
  \institution{National University of Singapore}
  \city{Singapore}
  \country{Singapore}}
\email{E0675868@u.nus.edu}

\author{Dinil Mon Divakaran}
\affiliation{%
  \institution{ A*STAR Institute of Advanced Intelligence and Computing (IAIC)}
    \city{Singapore}
  \country{Singapore}
  }

\author{Mohan Gurusamy}
\affiliation{%
  \institution{National University of Singapore}
  \city{Singapore}
  \country{Singapore}}


\renewcommand{\shortauthors}{B. Wu et al.}


\begin{abstract}

Early-stage encrypted traffic classifiers make decisions from only the first few packets of a connection, enabling low-latency classification in high-speed networks. 
Although these classifiers often achieve high accuracy on clean traffic, their robustness under realistic traffic morphing remains unclear. 
Existing adversarial traffic generation methods typically assume access to a complete flow, rely on feature-level perturbations that may not map to valid packet sequences, or require iterative runtime feedback. 
These assumptions do not fully match early-stage morphing. 
Attacks are confined to a short and structurally rigid initial packet sequence, and morphing must commit actions within these constraints. 

We present \textsc{NetGuise}, a target-conditioned traffic morphing framework for measuring and exploiting the \emph{morphability} of early-stage classifiers under realistic shaping constraints. 
\textsc{NetGuise} generates packet-level tunnel-layer padding and non-negative delay actions that preserve packet direction, ordering, and packet count while respecting MTU, bandwidth, and latency constraints. 
We instantiate \textsc{NetGuise} in the use case of \emph{censorship evasion}, where blocked traffic must be morphed to resemble a specific allowed application.
We evaluate \textsc{NetGuise} under two complementary observation interfaces.
The \emph{history-only causal interface} observes only the packet history available so far and commits one shaping action at a time, supporting online packet-by-packet execution.
The \emph{full-window oracle interface} provides the complete early-stage window before action generation and serves as our main controlled upper-bound characterization of early-stage morphability across datasets, classifiers, and source--target pairs.
The gap between the two interfaces quantifies the additional difficulty introduced by causal execution.
On VANT, \textsc{NetGuise} achieves 33.2\% mean targeted evasion under the history-only causal interface, compared with 82.0\% under the full-window oracle interface; on UTMobileNet, the corresponding results are 5.6\% and 81.8\%.
Despite this reduction, \textsc{NetGuise} consistently outperforms the strongest causal baseline, BAP, achieving 33.2\% versus 15.48\% mean targeted evasion on VANT and 5.6\% versus 4.11\% on UTMobileNet.
On VANT, \textsc{NetGuise} also reduces bandwidth overhead from 19.80\% to 3.02\% and generation latency from 37.42~ms/flow to 1.16~ms/flow relative to BAP, while requiring no inference-time classifier queries.
Under the full-window oracle interface, \textsc{NetGuise} achieves up to 85.7\% targeted evasion and improves over the strongest baseline by up to 20.3\%.
It retains 43.5\% targeted evasion after four rounds of adaptive censor retraining and maintains 48\% targeted evasion under zero-shot cross-environment transfer.
Additional defense experiments show that extending the observation horizon and restricting directly manipulable features can substantially reduce evasion, while Mininet load-stress experiments show loss-free operation below the evaluated bottleneck capacity and increasing packet loss after the post-morphing demand exceeds that capacity.
These results show that early-stage classifiers can remain morphable despite high clean accuracy, and that robustness evaluation should account for targeted morphability, observation constraints, realistic shaping constraints, defender adaptation, and operational network conditions.

\end{abstract}






\authorsaddresses{%
This is a preprint of our paper accepted to
\textit{Proceedings of the ACM on Measurement and Analysis of Computing Systems (POMACS)},
December 2026, and to be presented at ACM SIGMETRICS 2027.
Corresponding author: Binghui Wu
(\href{mailto:E0675868@u.nus.edu}{E0675868@u.nus.edu}).
Authors' Contact Information:
Binghui Wu and Mohan Gurusamy, National University of Singapore;
Dinil Mon Divakaran, A*STAR Institute of Advanced Intelligence and Computing (IAIC), Singapore.
}

\maketitle

\section{Introduction}\label{sec:intro}

Network-traffic classifiers are increasingly used to support security, monitoring, and network-management tasks, including anomaly detection, website and IoT device fingerprinting, censorship enforcement, traffic analysis, and performance diagnosis~\cite{Usenix-2023-website-fingerprinting, ndss-2021flowlens,asiaccs-2022-censor, Traffic-classification-2025-KDD,  ZEST-2024,2024-netdiffusion,blackbox-2025-arXiv-Kexu}. 
As encrypted protocols such as TLS~\cite{TLS-1.3} become widely deployed, packet payloads are no longer available for deep packet inspection. 
Modern machine learning (ML)-based traffic classifiers (MTCs) for encrypted traffic have emerged as a dominant paradigm, operating on traffic metadata alone (e.g., packet sizes, directions, and inter-arrival times)~\cite{ndss-2018-kitsune,NDSS-2020-flowprint, ccs-2021-nta, kexu-ndss-2023-detecting,Usenix-2023-website-fingerprinting}.

At the same time, large networks further push MTCs toward making early decisions.
ISPs, enterprise networks, and programmable data planes~\cite{p4-original-2014,p4-TBS-2021} must handle large numbers of concurrent flows, making it costly to buffer long flow histories before classification~\cite{2025-fastflow}.
Such in-network deployment motivates \textbf{early-stage} MTCs, where the classifier makes a decision from only the first few packets of a connection, i.e., an \textit{early packet sequence}, to reduce latency and processing cost~\cite{2025-fastflow,GGFAST-SIGCOMM-2023,ccs-2024-early-stage-wfp,JITI-2025-ACM}.
{This early-decision requirement is also reflected in 
measurements of real-world censorship infrastructure showing that encrypted traffic can be passively detected and blocked in real time~\cite{wu_HowGreatFirewall_2023}, while studies of early traffic classification optimize for the number of initial packets that must be observed before a reliable decision is made~\cite{2025-fastflow}. 
More generally, delaying classification requires longer per-flow observation and state, increasing decision latency and cost at high-throughput gateways~\cite{ML-Classificiation-NSDI-2025-Feamster}.
Prior works~\cite{early-stage-2007,Grad-BP-2010,GGFAST-SIGCOMM-2023,ccs-2024-early-stage-wfp,2025-fastflow,UniNet-2025,JITI-2025-ACM} have shown that early packet windows can be highly informative for accurate and low-latency traffic classification. 
Therefore, first-$N$-packet classification represents a practical trade-off between classification accuracy, decision latency, and per-flow resource cost.}

Despite growing interest in early-stage MTCs, their robustness against adversarial evasion under realistic network constraints remains insufficiently tested. 
This problem is non-trivial because early-stage classification imposes constraints that are absent, or much weaker, in full-flow settings.
\GapOne\textit{Limited early-stage action space}. 
Because the classifier observes only an early-stage window, the evader must perturb a short and structurally rigid part of the connection. 
In this window, packet directions, handshake behavior~\cite{TLS-1.3}, and early timing patterns leave limited room for realizable and semantics-preserving modification. 
As a result, perturbations that may be feasible over a complete flow can become infeasible within the first few packets.
\GapTwo\textit{Realizability and perturbation asymmetry}. 
MTCs operate on extracted features, but the evader must produce a concrete packet sequence that can traverse real network stacks~\cite{2024-netdiffusion}. 
Perturbing features in an abstract space does not guarantee that the result maps back to a valid packet sequence. 
Violations of TLS handshakes or application-layer semantics can make an adversarial example unrealizable~\cite{PANTS-2025-Usenix}. 
Moreover, realistic traffic shaping is asymmetric. 
Although packet sizes can be increased through padding, reducing payload size can change transport-layer behavior and expose the evasion attempt. 
Similarly, packet delays can be increased, but packets cannot be sent earlier than they arrive. 
In practice, safe shaping operations are therefore limited to additive padding and non-negative delay~\cite{2009-traffic-morphing}. 
This rules out the symmetric perturbation budgets assumed by many adversarial ML methods~\cite{2021-CCS-whitebox}.
\GapThree\textit{History-only causal execution}. 
A stricter challenge arises when future packet information is unavailable to the evader. 
At packet step $t$, the evader can use only the packet history observed so far and must commit the current shaping action before later packets are available. 
Once an action is committed, the corresponding packet cannot be modified again.

Existing adversarial traffic generation methods do not fully satisfy the requirements of early-stage robustness evaluation. 
Many prior methods~\cite{PANTS-2025-Usenix,Amoeba-2023-CoNext,blackbox-2025-arXiv-Kexu}  assume full-flow access, offline manipulation, feature-space perturbations, or iterative runtime feedback.
Some attacks~\cite{Amoeba-2023-CoNext,PANTS-2025-Usenix,blackbox-2025-arXiv-Kexu,2021-CCS-whitebox} optimize feature-level or flow-level representations rather than concrete packet-level actions. 
Such perturbations may appear effective against a classifier, but they fail to map back to valid packet sequences, exceed the adversary's permitted operations, or alter the original flow semantics~\cite{crawford2026adversarial,PANTS-2025-Usenix}. 
Solver-based methods can enforce richer constraints, but they typically require white-box access or expensive per-flow solving~\cite{PANTS-2025-Usenix,2009-traffic-morphing,2021-CCS-whitebox,whitebox-boosting-2024-expert}. 
Learning-based and mutation-based methods face a different difficulty. 
Sparse rewards, restricted actions, and abstract perturbation targets make it hard to learn stable targeted packet-level strategies with bounded overhead~\cite{blackbox-2025-arXiv-Kexu,BAP-2021-Usenix,Amoeba-2023-CoNext}. 
Moreover, methods that rely on runtime classifier queries can be expensive and impractical when decisions must be made over short packet windows~\cite{2025-fastflow}. 
These limitations mean that existing evasion methods do not directly provide a controlled robustness-evaluation framework for early-stage MTCs.
They highlight the need for an early-stage evader that generates concrete packet-level actions, remains realizable under a specified threat model, supports full-window oracle evaluation, and can be tested under history-only causal execution without runtime queries to the deployed classifier.

To address these gaps, we present \sol, a constraint-aware adversarial traffic morphing framework for evaluating the robustness of early-stage encrypted traffic classifiers under realistic shaping constraints.
We evaluate \sol in the censorship setting, while its architectural design is applicable to other metadata-based MTCs.
Given a source early-stage window and a specified target class, \sol generates concrete packet-level actions that morph the source window toward the target class. 
Rather than relying on per-flow solvers or runtime classifier queries, \sol learns a reusable morpher with local surrogate classifiers and transfers it to unseen classifiers in a no-query setting.
\sol consists of four components. 
i)~First, a morphability-aware target-reference selection strategy accounts for the asymmetry of additive-only morphing and selects a compatible reference within the specified target class. 
ii)~Second, a sequence encoder maps the source window and selected target reference into a shared packet-level representation space. 
iii)~Third, a target-conditioned Transformer aligns the source and target representations and decodes packet-level padding and delay actions. 
iv)~Fourth, a constraint projection layer maps each raw action into the feasible shaping space, enforcing additive-only padding, non-negative delay, MTU compliance, direction and packet-count preservation, and configured bandwidth and latency budgets.
The target class is fixed by the evaluation objective, and \sol selects only a compatible reference within that class. 
Some references can be reached by adding packet size or delay, whereas others would require shrinking packets or changing directions and are infeasible under the threat model.
The design of \sol separates the robustness-evaluation objective from the observation interface. 
{We consider two complementary interfaces with different purposes.
Under the \emph{history-only causal interface} ($\mathcal{I}^\text{causal}$), the morpher observes only the packet history available up to the current time and must commit each shaping action before further packets are observed. 
This interface supports online packet-by-packet execution.
For our controlled characterization, we define a \emph{full-window oracle interface} ($\mathcal{I}^\text{oracle}$), in which the complete early-stage window is available before actions are generated. 
Because the action for an early packet may depend on later packets in the same window, we treat this offline setting as an upper bound on early-stage morphability. 
The oracle interface provides a controlled setting for systematic comparison across datasets, classifier architectures, baselines, and source--target directions under the same observation and shaping constraints.}

Although \sol is agnostic to the deployment context, we instantiate and evaluate it in the use case of \textbf{censorship evasion}. 
This use case is particularly demanding for early-stage morphability evaluation. 
First, censorship evasion has direct societal relevance because state-level censorship affects access to information and privacy-enhancing technologies for large populations~\cite{NDSS-2025-censorship-process,ccs-2025-censorship-for-DPI,NDSS-2024-censorship-process}. 
Second, early-stage classification is operationally important for censors operating at ISP or backbone scale: a censor must classify and act before substantive application data is exchanged, both to enforce blocking and to limit collateral damage~\cite{wu_HowGreatFirewall_2023,NDSS-2024-censorship-process,IMC-2017-censorship, bocovich_SnowflakeCensorshipCircumvention_2024, wrana_SoKSpectreSurveillance_2025}. 
Third, censorship evasion induces a strict targeted objective. 
Unlike anomaly detection, where evasion often means being classified as benign, or website fingerprinting, where landing in a broad unmonitored set may suffice, censorship evasion requires blocked traffic to be classified as a \emph{specific allowed application}~\cite{wu_HowGreatFirewall_2023}. 
This makes it possible to measure which source--target directions are morphable and why.

{We evaluate early-stage morphability in two complementary stages.
We first establish how morphable these classifiers can be under the controlled oracle setting $\mathcal{I}^{\text{oracle}}$, and then examine how much of this capability remains under causal online execution $\mathcal{I}^{\text{causal}}$.
}
Under {$\mathcal{I}^{\text{oracle}}$}, our comprehensive evaluation spans three datasets, four censor architectures, and both targeted and untargeted evasion objectives.
{This setting provides a controlled upper bound on early-stage morphability rather than an online deployment model.}
\sol achieves up to 85.7\% targeted evasion while satisfying all specified shaping constraints under our two-ended tunnel-shaping threat model. 
It improves targeted evasion over the strongest baseline by up to 20.3\% under the same early-stage constraints. 
In a four-round adaptive arms-race evaluation, where both the evader and censor retrain after each round, \sol retains 43.5\% targeted evasion, compared with 17.8\% for the strongest baseline. 
Under zero-shot cross-environment transfer, where the morpher has no access to traffic from the target environment, \sol maintains 48\% targeted evasion and over 63\% untargeted evasion. 
{
We then move to $\mathcal{I}^{\text{causal}}$, where future packet positions are hidden and each shaping action must be committed before later packets are observed, supporting online packet-by-packet execution.
Under this stricter interface, \sol achieves 33.2\% mean T-ESR on VANT and 5.6\% on UTMobileNet.
Despite the substantial reduction from the oracle setting, \sol consistently outperforms the strongest causal baseline, BAP, achieving 33.2\% versus 15.48\% mean targeted evasion on VANT and 5.6\% versus 4.11\% on UTMobileNet.
}
On VANT, \sol further reduces bandwidth overhead to 3.02\% and generation latency to 1.16~ms/flow while requiring no inference-time classifier queries.
Across our validity checks, generated flows preserve packet direction, ordering, and packet count, while respecting MTU, bandwidth, and latency constraints. 
This paper makes the following contributions:

\begin{enumerate}

\item We formulate \textbf{early-stage traffic morphability} as a robustness-evaluation problem for encrypted early-stage traffic classifiers. 
To the best of our knowledge, this is the first work to systematically evaluate the robustness of early-stage MTCs under realistic packet-level shaping constraints.

\item We design \sol, a target-conditioned traffic morphing framework that generates concrete packet-level padding and delay actions. 
The generated actions preserve packet direction, ordering, and packet count, while satisfying MTU, bandwidth, and latency constraints.

\item We introduce a morphability-aware target-reference selection and alignment strategy that accounts for source--target structural compatibility under additive-only shaping. 
The target class is fixed by the evaluation objective. 
\sol selects a compatible reference only within that specified class.

\item We evaluate \sol across three datasets and four classifier architectures. 
Our main controlled characterization uses the {full-window oracle interface ($\mathcal{I}^{\text{oracle}}$)}, while the {history-only causal interface ($\mathcal{I}^{\text{causal}}$)} evaluates online packet-by-packet execution. 
{We further study adaptive and feature-restricted defenders, extended observation windows, and end-to-end validity.}
The results show that early-stage classifiers can remain highly morphable despite high clean accuracy under the oracle upper bound, while causal morphing is substantially more challenging and environment dependent. 
{They also show that robustness depends jointly on source--target morphability, observation constraints, defender adaptation, and operational conditions.}
To support reproducibility, we release the source code, training pipeline, testbed, and datasets\footnote{\url{https://github.com/iwantto321/NetGuise}}.


\end{enumerate}

\noindent\textbf{Ethics considerations.}
This work studies dual-use traffic-morphing techniques to expose limitations in passive traffic analysis; all evaluations use public or controlled datasets containing no human-subject data, personally identifiable information, or plaintext payloads, and we focus on structural robustness analysis rather than deployment-ready evasion procedures.

\section{Related Work} \label{sec:related_work}

\noindent\textbf{Early-Stage Traffic Classification.}
For effective traffic control at carrier-grade network speeds, censorship systems deployed at ISP- or national-scale gateways must classify connections from only the first few packets~\cite{2025-fastflow,wu_HowGreatFirewall_2023,GGFAST-SIGCOMM-2023,ML-Classificiation-NSDI-2025-Feamster}.
This has led to a class of \emph{early-stage} MTCs that issue blocking decisions based on a fixed packet budget or observation window~\cite{UniNet-2025,ccs-2021-nta, ZEST-2024,2025-fastflow,GGFAST-SIGCOMM-2023,Grad-BP-2010,flowpic-tnsm-2021,ton-2022,ML-Classificiation-NSDI-2025-Feamster,Francesco-dataset-2019,ML-2021-IMC,CCS-2021-automation-analysis}.
Early approaches rely on statistical features or fixed windows. 
With programmable switches~\cite{p4-original-2014}, lightweight or hybrid ML logic can now be deployed directly in the network fabric at Tbps line rates (e.g.,~\cite{ccs-2021-nta,Unsenix2023-IDP-programmable-switch-flow,p4-TBS-2021}). Such deployments inherently favor early-stage inspection, as buffering long flows or performing complex analysis is infeasible at line rate~\cite{suupport-p4idea-2025-acm-com-survey}.
Among the early-stage classifiers, GGFAST~\cite{GGFAST-SIGCOMM-2023} constructs n-gram fingerprints from the first 50 packets, FlowPic~\cite{flowpic-tnsm-2021} maps traffic into time-binned images for CNN classification, and Grad-BP~\cite{Grad-BP-2010} applies SVMs over cumulative statistics.
Recent SOTA systems leverage deep neural architectures to extract temporal structure from early-stage windows. FastFlow~\cite{2025-fastflow} employs adaptive LSTMs, while UniNet~\cite{UniNet-2025} uses Transformer-based self-attention to model long-range dependencies within packet sequences.
By concentrating discriminative power within the initial packets, these models create an early-stage dilemma: evasion must occur within the most constrained and semantically rigid portion of the flow (e.g., TCP/TLS handshakes). This constraint motivates designs that can operate within protocol semantics while disrupting early-stage classifiers.

\noindent\textbf{Adversarial Network Evasion.}
Traffic morphing seeks to reshape the observable characteristics of a flow to evade classification~\cite{2009-traffic-morphing}. Modern attacks~\cite{PANTS-2025-Usenix, blackbox-2025-arXiv-Kexu,Amoeba-2023-CoNext,BAP-2021-Usenix,2021-CCS-whitebox, whitebox-2019,whitebox-JASC-2021,gong2022surakav,2022-TDSC-Whitebox,blackbox-2019-ccs-Geneva,blackbox-2021-eswa,IDSGAN-2022-blackbox,ipets-2023,holland2024detorrent,evasionformalware-2025,Certa-2025-Usenix-Security} fall into white-box and black-box paradigms, both with limitations against a realistic censor. 
White-box methods use gradient information to induce misclassification, but weakly constrained perturbations may violate flow semantics.
BAP~\cite{BAP-2021-Usenix} generates input-independent perturbations for live
traffic and uses remapping and regularization to enforce traffic-feature constraints.
PANTS~\cite{PANTS-2025-Usenix} enforces realizability by coupling gradients with SMT solvers; however, it requires white-box access and introduces a trade-off, as formal solving is computationally expensive and often fails (\texttt{UNSAT}) under realistic network jitter.
Black-box approaches, such as Amoeba~\cite{Amoeba-2023-CoNext}, rely on RL to discover evasion policies. These methods suffer from a semantic gap: optimization is driven solely by misclassification reward, leading to inefficient or anomalous perturbations. Moreover, iterative RL inference incurs prohibitive latency ($>1$\,s), limiting scalability~\cite{Amoeba-2023-CoNext}. 

\noindent\textbf{Network Traffic Generation.}
A parallel line of work applies generative models or simulations to synthesize network traffic~\cite{2024-netdiffusion,NetShare-2022-SIGCOMM, ns-3-2008,yans-2006,Traffic-generation-GAN-IMC-2020,DYNAMO-2022}.
Generative systems such as DoppelGANger~\cite{Traffic-generation-GAN-IMC-2020}, NetShare~\cite{NetShare-2022-SIGCOMM}, and NetDiffusion~\cite{2024-netdiffusion} employ GANs and diffusion models to generate high-fidelity synthetic headers for privacy-preserving analysis.
However, these approaches are not well suited for censorship evasion.
They exhibit an objective mismatch: generative models optimize for distributional fidelity, whereas evasion requires deviation from learned decision boundaries.
Moreover, they operate in an abstract feature space~\cite{feature-space-pertubation-2020-sp} and do not solve the inverse mapping to discrete, valid packet sequences~\cite{realistic-blackbox-2023}.

\section{Threat Model}
\label{sec:threat_model}

\begin{figure}[htbp] 
    \centering
    \includegraphics[width=0.85\linewidth]{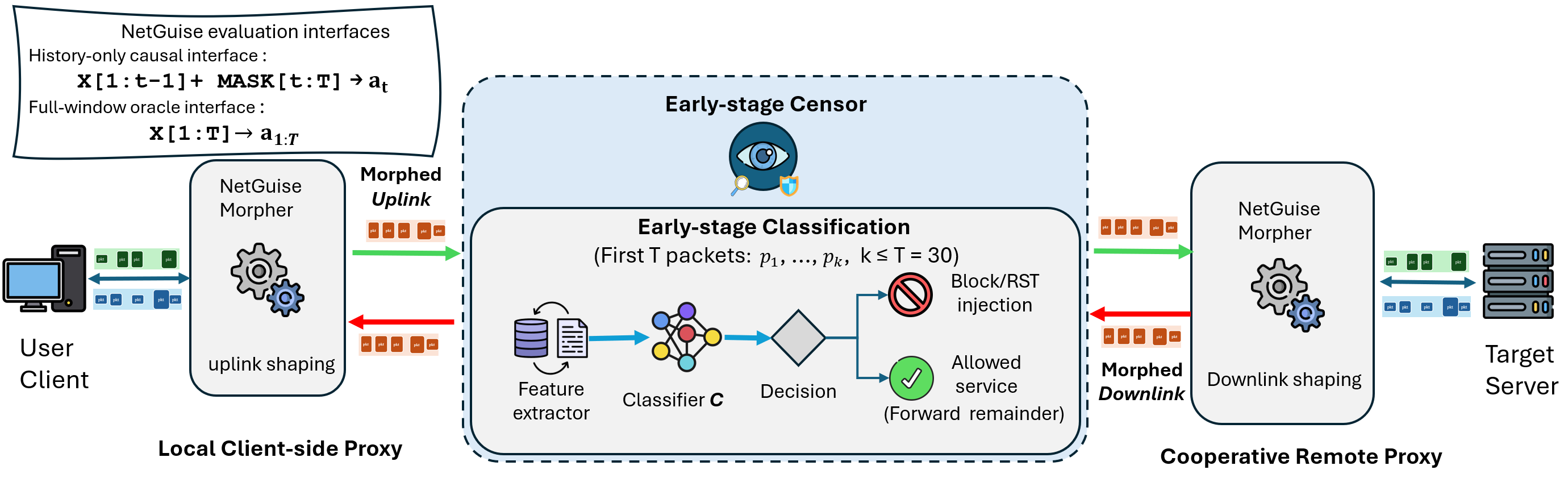}
    \Description{Diagram showing the threat model and execution modes supported by the system.}
    \caption{Threat model and evaluation interfaces.
The history-only causal interface ($\mathcal{I}^{\text{causal}}$) observes only $X_{1:t}$ and commits the current action $a_t$ before future packets are available, supporting online packet-by-packet execution. The {full-window oracle interface ($\mathcal{I}^{\text{oracle}}$)} maps
$X_{1:T}$ to $a_{1:T}$ and serves as a controlled upper bound on early-stage
morphability rather than an online deployment model.}
    \label{fig:threatmodel}
\end{figure}

We formulate adversarial traffic morphing in the context of early-stage encrypted traffic classification for censorship evasion. 
The censor is an on-path classifier that observes bidirectional packet metadata and makes a blocking decision from an early packet sequence of a connection. 
The `adversary' aims to morph traffic from a source application into traffic that is classified as an allowed target application, while preserving the original application semantics and keeping bandwidth and latency overhead bounded.
We assume a two-ended tunnel-shaping threat model consisting of a client-side morphing proxy and a cooperative relay located beyond the censor. 
The censor lies on the path between these two endpoints. 
The client-side proxy shapes client-to-relay packets before they reach the censor, while the relay shapes relay-to-client packets before they traverse the censor in the reverse direction. 
Thus, both traffic directions can be shaped, but each packet is shaped only by its transmitting endpoint. 
This deployment is consistent with practical proxy and pluggable-transport systems. 
Fig.~\ref{fig:threatmodel} summarizes the system model.

\noindent\textbf{Early-stage censor.} 
The censor is a stateful middlebox deployed at an ISP or network gateway. 
It observes packet metadata from both directions of a connection, including packet sizes, directions, and inter-arrival times, but not encrypted payload contents. 
The censor merges the two directions into a bidirectional sequence ordered by packet arrival time and classifies the connection using an early packet sequence rather than the completed flow. 
Unless otherwise stated, we use the first $T=30$ packets as the maximum early-stage observation window. 

\noindent\textbf{Adversarial objective.} 
Given a source flow with label $y_{\mathrm{src}}$, the adversary seeks to produce a morphed flow that the censor classifies as a selected allowed target class $y_{\mathrm{tgt}}\neq y_{\mathrm{src}}$. 
This is a targeted impersonation objective and is stricter than untargeted misclassification. 
At observation length $k$, targeted evasion succeeds if the censor predicts $y_{\mathrm{tgt}}$ from the morphed early-stage window $\tilde{x}_{1:k}$. 
We also report untargeted evasion, where success only requires the morphed early-stage window to be classified as any class other than $y_{\mathrm{src}}$.

\noindent\textbf{Adversary knowledge.} 
We adopt a no-query black-box transfer setting. 
The adversary trains local surrogate classifiers using independently collected traffic. 
The adversary has no access to the deployed censor's parameters, gradients, logits, confidence scores, training data, or query interface. 
\sol is optimized offline against local surrogates and then transferred to the deployed censor. 
During inference, \sol is query-free. 
Under the {full-window oracle interface ($\mathcal{I}^{\text{oracle}}$)}, it generates actions for the complete early packet sequence in one forward pass. 
{This interface assumes access to the complete early-stage window before action generation and is used as a controlled oracle upper bound rather than an online deployment setting.}
Under the {history-only causal interface ($\mathcal{I}^{\text{causal}}$)}, it makes one causal action decision when a tunnel endpoint transmits a packet, using only the packet history observed so far {and without access to future packets}. 
If a baseline requires classifier queries at runtime, we report this explicitly as a stronger deployment assumption.

\noindent\textbf{Permitted operations.}
The adversary may only apply tunnel-layer traffic shaping. 
For each packet, \sol may add tunnel-layer padding and non-negative transmission delay, but it cannot shrink packets, delete packets, inject packets, split packets, reorder packets, redirect packets, or change packet directions. 
Packet ordering, packet count, packet direction, and the original application payload are preserved. 
Generated packets must satisfy MTU constraints and configured bandwidth and latency budgets. 
Unless otherwise stated, the default bandwidth and latency overhead budgets are 30\% and 50\%, respectively.

\section{Architecture and Methodology}
\label{sec:design}

\sol is a target-conditioned traffic morphing framework for early-stage classifier evasion. 
Given a source packet sequence $X=(x_1,\ldots,x_T)$ and a target class $y_{\mathrm{tgt}}$, \sol generates padding and delay actions that steer the morphed sequence toward the target class while preserving packet direction, ordering, and packet count. 
The generated actions also satisfy the specified constraints. 
We evaluate the same architecture under two complementary interfaces.
{The full-window oracle interface is used for our main controlled characterization of morphability, while the history-only causal interface supports online packet-by-packet execution.}

\subsection{Problem Formulation}
\label{sec:formulation}

Let each packet be represented as $x_t=(s_t,d_t,\tau_t)$, where $s_t$ is packet size, $d_t$ is direction, and $\tau_t$ is inter-arrival time.
An early-stage window is denoted as $X=(x_1,\ldots,x_T)$, where $T=30$ is the default early-stage observation window unless otherwise stated.
Given a source early-stage window $X$ with source label $y_{\mathrm{src}}$, the adversary selects a target class $y_{\mathrm{tgt}}\neq y_{\mathrm{src}}$ and seeks a morphed window $\tilde{X}$ such that an early-stage classifier predicts $y_{\mathrm{tgt}}$ from $\tilde{X}_{1:T}$.

Because the deployed censor is a black box, \sol trains against a local differentiable surrogate classifier $\mathcal{S}$ and transfers the learned morpher to the deployed classifier.
The generator $G_\theta$ is optimized offline to minimize the targeted classification loss on the default early-stage window:
\begin{equation}
\label{eq:adv_loss_basic}
    \mathcal{L}_{\mathrm{adv}}
    =
    CE\!\left(
        \mathcal{S}(\tilde{X}_{1:T}),
        y_{\mathrm{tgt}}
    \right).
\end{equation}
The surrogate parameters are frozen when training the morpher; gradients update only $G_\theta$.
For the {history-only causal evaluation}, the same classifier can be evaluated at earlier decision points $k \leq T$, while the {full-window oracle interface} uses $T=30$.

\subsection{\sol Architecture}
\label{sec:architecture}

\begin{figure*}[htbp]
    \centering
    \includegraphics[width=0.8\linewidth]{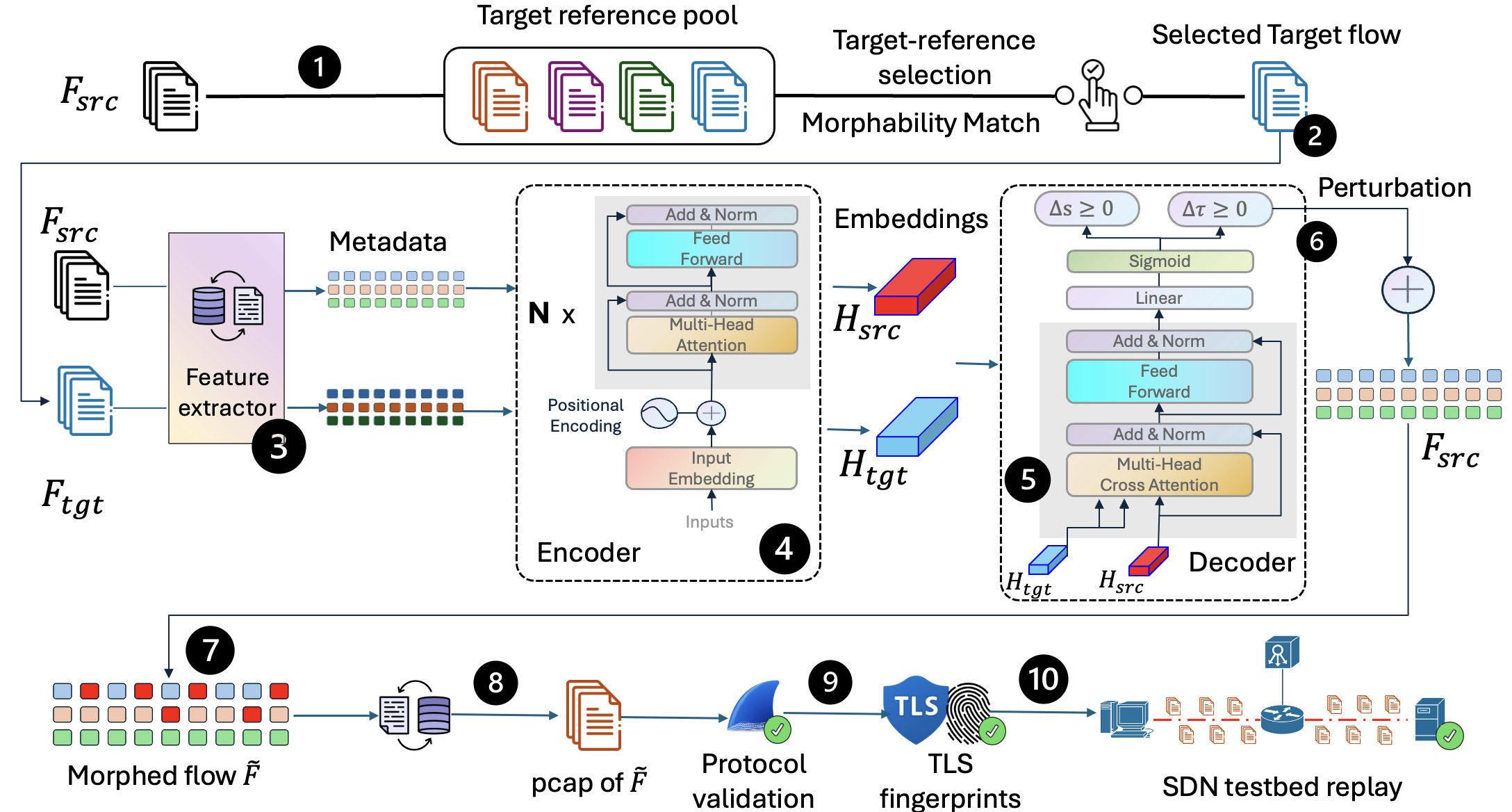}
    \caption{Overview of \sol: a target-conditioned pipeline that selects a compatible target reference, generates packet-level padding and delay actions, projects them into feasible shaping constraints, and validates the resulting morphed trace.}
    
    \label{fig:model_arch}
\end{figure*}

\sol learns a parameterized generator $G_\theta$ that amortizes constraint-aware morphing into a reusable conditional model.
Given a source early-stage window and a target condition, the generator produces packet-level shaping actions that are subsequently projected into the feasible tunnel-shaping space.
Fig.~\ref{fig:model_arch} illustrates the end-to-end pipeline.
Steps~1--6 comprise the four architectural components: morphability-aware target selection, sequence encoding, target-conditioned alignment, and action decoding with constraint projection.
Steps~7--10 perform validity assessment, including packet-trace reconstruction, protocol checks, TLS fingerprint inspection, and SDN testbed replay.

\subsubsection{Morphability-Aware Target Selection (Steps~1--2 in Fig.~\ref{fig:model_arch})}
\label{sec:target-selection}
The target class $y_{\mathrm{tgt}}$ is given by the attack objective.
\sol therefore selects a structurally compatible target reference from the given target class. 
This reference is used to condition the morpher and guide the source early-stage window toward the selected target class.
Under additive-only shaping, morphability is asymmetric; padding and delay can increase packet sizes and inter-arrival times, but they cannot shrink packets, advance packets in time, change packet directions, or modify packet count. 
We therefore rank candidate target references using a morphability score:
\begin{equation}
\label{eq:morphability_score}
    S(X_{\mathrm{src}},X_{\mathrm{tgt}})
    =
    \alpha C_s(X_{\mathrm{src}},X_{\mathrm{tgt}})
    +
    \beta C_\tau(X_{\mathrm{src}},X_{\mathrm{tgt}})
    +
    \gamma C_d(X_{\mathrm{src}},X_{\mathrm{tgt}}),
\end{equation}
where $C_s$, $C_\tau$, and $C_d$ measure size, timing, and direction compatibility, respectively, and $\alpha+\beta+\gamma=1$. 
Size compatibility favors references reachable by feasible padding, timing compatibility favors references reachable by feasible delay, and direction compatibility favors references whose packet-direction sequence already matches the source. 
We set $(\alpha,\beta,\gamma)=(0.3,0.3,0.4)$ throughout our experiments.
The detailed definitions and a sensitivity analysis of the weighting coefficients are provided in Appendix~\ref{app:morphability_score}.

For each source early-stage window $X_{\mathrm{src}}$ and fixed target class $y_{\mathrm{tgt}}$, we sample a candidate set
$\mathcal{C}_{y_{\mathrm{tgt}}}$ of size $K$ from the evader's training pool $\mathcal{P}_{y_{\mathrm{tgt}}}$ and select the highest-scoring reference:
\begin{equation}
\label{eq:target_select}
    X_{\mathrm{ref}}
    =
    \arg\max_{X_k\in\mathcal{C}_{y_{\mathrm{tgt}}}}
    S(X_{\mathrm{src}},X_k).
\end{equation}
This selection uses only evader-side training data and does not use deployed-censor outputs, gradients, confidence scores, runtime queries, or test-time attack success. 
{Appendix~\ref{sec:ablation_target_selection} further evaluates the target-selection strategy, including a distance-matched control that separates overall source--reference proximity from asymmetric shaping feasibility.}

\subsubsection{Sequence Encoding (Steps~3--4 in Fig.~\ref{fig:model_arch})}

The source early-stage window is represented as a sequence of packet features $\{(s_t,d_t,\tau_t)\}_{t=1}^{T}$.
A packet embedding layer and positional encoding map this sequence into a hidden representation $H_{\mathrm{src}}\in\mathbb{R}^{T\times d}$.
The target condition, either a reference target window or a target-class prototype, is encoded into $H_{\mathrm{tgt}}\in\mathbb{R}^{T\times d}$.
The same encoder parameters are shared between source and target representations, so both are embedded into a common packet-level representation space.

\subsubsection{Target-Conditioned Alignment (Step~5 in Fig.~\ref{fig:model_arch})}

To align source and target structure, \sol uses target-conditioned cross-attention.
The source hidden states act as queries, while the target representation provides keys and values:
\begin{equation}
\label{eq:attention}
    Z
    =
    H_{\mathrm{src}}
    +
    \mathrm{Attention}
    \left(
        Q=H_{\mathrm{src}},
        K=H_{\mathrm{tgt}},
        V=H_{\mathrm{tgt}}
    \right).
\end{equation}
The residual connection preserves the source sequence structure, while the attention output injects target-conditioned information indicating how the source should be morphed.
This design lets the morpher learn where packet-level shaping is useful without requiring runtime queries to the deployed censor. Appendix~\ref{sec:ablation_ca} ablates this design against other conditional baselines.

\subsubsection{Action Decoding (Step~6 in Fig.~\ref{fig:model_arch})}

A lightweight decoder maps each aligned hidden state $Z_t$ to raw action
logits:
\begin{equation}
\label{eq:decoder}
    o_t
    =
    (o^s_t,o^\tau_t)
    =
    \mathrm{MLP}(Z_t),
\end{equation}
where $o^s_t$ controls padding and $o^\tau_t$ controls delay.
These raw outputs are not directly applied to the traffic.
They are passed through the constraint projection layer described next, which maps them into feasible tunnel-layer shaping actions.

\subsection{Tunnel-Shaping Action Projection}
\label{sec:action_projection}

\sol only permits additive padding and non-negative transmission delay.
For each packet, the final action is
\[
    a_t=(\Delta s_t,\Delta \tau_t),
\]
and the morphed packet metadata are
\[
    \tilde{s}_t=s_t+\Delta s_t,
    \qquad
    \tilde{\tau}_t=\tau_t+\Delta\tau_t.
\]
Packet direction $d_t$ is preserved unchanged.
To enforce the permitted action space, \sol projects decoder outputs into the feasible tunnel-shaping region.
Let $M$ denote the MTU, and let $B_s^{(t)}$ and $B_\tau^{(t)}$ denote the remaining bandwidth and latency budgets before packet $t$ is shaped. 
In deployment, $M$ should be configured according to the effective tunnel PMTU after accounting for encapsulation overhead, so that the resulting tunneled packet does not require fragmentation~\cite{rfc4459}.
The projected action is
\begin{equation}
\label{eq:size_projection}
    \Delta s_t
    =
    \sigma(o^s_t)
    \cdot
    \min\{M-s_t,\; B_s^{(t)}\},
\end{equation}
\begin{equation}
\label{eq:delay_projection}
    \Delta \tau_t
    =
    \sigma(o^\tau_t)
    \cdot
    B_\tau^{(t)},
\end{equation}
where $\sigma(\cdot)$ maps raw logits to $[0,1]$. 
Because Eqs.~\ref{eq:size_projection}--\ref{eq:delay_projection} consume the remaining budgets sequentially, earlier packets may receive a larger share of the available padding or delay budget. We evaluate this effect using a full-window hindsight allocation that jointly redistributes the same total budgets across the packet window (Appendix~\ref{app:budget_allocation}).
This projection enforces
\[
    \Delta s_t\ge 0,
    \qquad
    \Delta \tau_t\ge 0,
    \qquad
    \tilde{s}_t\le M.
\]
After each action, the remaining budgets are updated as
\[
    B_s^{(t+1)}
    =
    B_s^{(t)}-\Delta s_t,
    \qquad
    B_\tau^{(t+1)}
    =
    B_\tau^{(t)}-\Delta \tau_t.
\]
Thus, the cumulative overheads satisfy
\[
    \sum_{t=1}^{T}\Delta s_t\le B_s,
    \qquad
    \sum_{t=1}^{T}\Delta \tau_t\le B_\tau.
\]
Because the action space is additive-only, \sol cannot shrink packets, change packet directions, reorder packets, inject packets, or delete packets.
The generated traffic is therefore realizable under the two-ended tunnel-shaping threat model in \S\ref{sec:threat_model}.

\subsection{Evaluation Interfaces}
\label{sec:execution_modes}

\sol is primarily evaluated under a {full-window oracle interface ($\mathcal{I}^{\text{oracle}}$)} similar to previous works~\cite{PANTS-2025-Usenix,BAP-2021-Usenix,Amoeba-2023-CoNext}. 
{$\mathcal{I}^{\text{oracle}}$ serves as our main controlled characterization of early-stage morphability rather than an online deployment setting.}
To quantify the additional difficulty of history-only execution, we also evaluate the same target-conditioned morpher and projection layer {under a history-only causal interface ($\mathcal{I}^{\text{causal}}$)}.
Unlike the full-window settings used by prior work, {$\mathcal{I}^{\text{causal}}$} removes access to future packets and enforces commit-once packet actions.

\noindent\textbf{{Full-window oracle interface ($\mathcal{I}^{\text{oracle}}$).}}
Under {$\mathcal{I}^{\text{oracle}}$}, \sol receives the complete early-stage window $X_{1:T}$ and generates actions for all $T$ packets:
\[
    G_\theta(X_{1:T},y_{\mathrm{tgt}}) = (a_1,\ldots,a_T).
\]
{$\mathcal{I}^{\text{oracle}}$} matches the observation setting of the primary baselines and provides a controlled comparison under the same early-stage window and shaping constraints.
{Because the action for an early packet may depend on later packets in the same window, a literal online implementation would need to withhold the complete early-stage window before releasing the first shaped packet. We therefore treat $\mathcal{I}^{\text{oracle}}$ as an upper bound on early-stage morphability and do not include such hypothetical window-collection delay in the shaping-latency budget.}

\noindent\textbf{{History-only causal interface ($\mathcal{I}^{\text{causal}}$).}}
Under {$\mathcal{I}^{\text{causal}}$}, at packet step $t$, the morpher observes only the packet history $x_{1:t}$ and replaces all unobserved future positions with a learned mask token:
\begin{equation}
\label{eq:masked_input}
    \bar{x}^{(t)}
    =
    [x_1,\ldots,x_t,\texttt{MASK},\ldots,\texttt{MASK}].
\end{equation}
The Transformer still processes a length-$T$ sequence, but future packet features are not visible. 
The decoder may output candidate actions for all positions, but \sol commits only the current action $a_t$ and discards the future outputs. 
Once committed, $a_t$ cannot be revised. 
{$\mathcal{I}^{\text{causal}}$ supports online packet-by-packet execution.}

\noindent\textbf{Use of generative AI.}
Generative AI tools were used to assist with language editing, presentation refinement, and manuscript organization.
They were not used to generate experimental results.

\section{Evaluation}
\label{sec:evaluation}

{
We evaluate two different settings.
i)~We first establish the controlled oracle characterization under $\mathcal{I}^{\text{oracle}}$, which serves as our main benchmark for measuring early-stage morphability across datasets, classifiers, and source--target pairs.
This setting assumes access to the complete early-stage window and is not intended to represent online deployment.
ii)~We then evaluate how much of this morphability remains under $\mathcal{I}^{\text{causal}}$, where future packet positions are hidden and each action must be committed as packets arrive, supporting online packet-by-packet execution.
}
{
Across these settings, we examine evasion effectiveness, traffic validity, robustness under adaptation, operational overhead, and generation latency.
}
Implementation details and hyperparameters are provided in Appendix~\ref{app:hparams}.
We organize the evaluation around five research questions:

\begin{itemize}

\item \textbf{RQ1 ({Oracle characterization of evasion effectiveness}).}
{How effective is \sol at bypassing early-stage traffic classifiers
under the no-query black-box transfer setting when evaluated under
$\mathcal{I}^{\text{oracle}}$, for both targeted and untargeted objectives?}

\item \textbf{RQ2 ({History-only causal execution}).}
{How much of this morphability remains under
$\mathcal{I}^{\text{causal}}$, where each packet action must be committed
before future packets are observed?}

\item \textbf{RQ3 (Robustness under defender adaptation).}
{How does \sol perform against adaptive defenses that retrain the
classifier, extend its observation horizon, or restrict features that can
be directly manipulated by traffic shaping?}

\item \textbf{RQ4 (Operational cost).}
{What overhead does \sol incur, and how does its implementation behave
as offered load approaches the bottleneck capacity?}

\item \textbf{RQ5 (Traffic validity).}
Do morphed flows remain realizable packet traces that preserve the
specified protocol and action-space constraints, or are they only
feature-space artifacts?

\end{itemize}

 \subsection{Datasets \& Evaluation Protocols}
\label{sec:datasets}

\noindent\textbf{Datasets.}
We evaluate \sol across three heterogeneous environments containing full packet captures (\texttt{pcap}). \textbf{UTMobileNet-2021}~\cite{UT-Mobile-2021}
captures high-jitter cellular mobile traffic across 12 application classes. \textbf{VANT-2024}~\cite{VANT-2024} provides encrypted VPN tunnel
traffic across 5 classes: C2, Streaming, Chat, VoIP, and File Transfer.
\textbf{VideoPF-2025} is a custom adaptive-bitrate video streaming dataset with 4 classes, collected under realistic network shaping following
established methodology~\cite{Francesco-dataset-2019}. 
Together, these datasets cover stable VPN tunnels, bursty mobile traffic, and adaptive video streams. Detailed dataset compositions are provided in Appendix~\ref{app:datasets}.

Each dataset is partitioned into disjoint, class-stratified subsets: 50\% for training the censor classifier, 30\% for training the evader including the local surrogate and morpher, and 20\% for evaluation. 
This split enforces information asymmetry: the evader does not train on the censor's training flows. Unless otherwise stated, attacks are evaluated within a maximum early-stage horizon of $T=30$ packets. 
For history-only causal experiments, we additionally report performance at observation lengths $k\in\{5,10,15,20,30\}$ to model censors that may decide at different points within the same early-stage window.
We use the VANT dataset as the primary dataset for adaptive-censor experiments (\S\ref{sec:adaptive_censor}) because it contains censorship-relevant classes such as C2 and VoIP. C2 traffic captures tunnel and proxy-like behavior targeted by systems such as the GFW~\cite{wu_HowGreatFirewall_2023} and TSPU~\cite{TSPU-2022}, while VoIP services are blocked in multiple jurisdictions~\cite{vafa_LearningCensoredExperiences_2025a}.

\subsection{Censor Models and Baselines}
\label{sec:classifiers}

\noindent\textbf{Censor models.}
We evaluate \sol against four classifier families spanning rule-based, statistical, and neural approaches: \textbf{GGFAST}~\cite{GGFAST-SIGCOMM-2023}, a payload-agnostic Naive Bayes classifier using packet length and direction $n$-gram features; 
\textbf{Grad-BP}~\cite{Grad-BP-2010}, an SVM-based classifier trained on aggregate flow statistics; \textbf{FastFlow}~\cite{2025-fastflow}, an LSTM model designed for early-stage classification; 
and \textbf{UniNet}~\cite{UniNet-2025}, a Transformer encoder that captures temporal dependencies across packet, flow, and session features. 
Surrogate models are restricted to FastFlow and UniNet, the differentiable architectures that support gradient-based morpher training. 
This setup allows us to evaluate transferability when the local surrogate $\mathcal{S}$ differs from the deployed censor $C$.

\noindent\textbf{Evasion baselines.}
We compare \sol with baselines representing major paradigms of adversarial traffic generation. 
\textbf{Random Padding (RandP)} applies random padding and delay under the same overhead budgets without classifier feedback. 
{\textbf{BAP}~\cite{BAP-2021-Usenix} learns blind adversarial perturbations offline against a differentiable model and can apply them to live traffic without access to future packets. 
For our evaluation, BAP is trained against the same local surrogate available to the evader and restricted to the same padding and non-negative-delay action space and overhead budgets as \sol. 
\textbf{Amoeba}~\cite{Amoeba-2023-CoNext} is a classifier-guided mutation attack that uses runtime feedback to select traffic modifications. 
\textbf{PANTS}~\cite{PANTS-2025-Usenix} is originally white-box and couples classifier information with constraint solving; we use a transfer-based black-box adaptation, denoted PANTS-BB, that generates adversarial flows against a local surrogate and evaluates them on unseen censors.
\sol, BAP, and PANTS-BB may use gradients from their local surrogate models during offline optimization where required, but none has access to gradients of the deployed censor. 
Amoeba does not require gradient access and instead relies on runtime classifier feedback.}

Under the full-window oracle interface ($\mathcal{I}^{\text{oracle}}$), all methods are evaluated over the same $T=30$ early-stage window and under the same shaping budgets.
\sol and PANTS-BB may use the complete early-stage window when generating their actions, while BAP retains its original blind perturbation mechanism and does not depend on future packets.
Amoeba operates on the complete early-stage window using classifier-guided mutation.
$\mathcal{I}^{\text{oracle}}$ therefore provides a controlled comparison of attack effectiveness under early-stage and action constraints.

Under the history-only causal interface ($\mathcal{I}^{\text{causal}}$), future packets are hidden and each action must be committed before later packets are observed.
BAP naturally satisfies this observation constraint because its learned perturbations are input-independent at runtime.
We evaluate BAP under the same action space, shaping budgets, and history-only information constraint as \sol.
We adapt Amoeba to $\mathcal{I}^{\text{causal}}$ so that it may modify only the currently observed packet using the observed history and remaining budgets, while retaining its per-packet classifier queries.
\sol operates under the same history-only, commit-once execution constraint but performs query-free inference from its trained morpher.
Thus, BAP provides the causal comparison under the same information budget, whereas causal Amoeba represents a stronger runtime-information setting through classifier feedback.

\subsection{Evaluation Metrics}
\label{sec:metrics}

\noindent\textbf{Evasion effectiveness.}
We report targeted and untargeted evasion success rates. Untargeted
evasion success rate (U-ESR) is the fraction of morphed flows classified
as any class other than the source class:
\begin{equation}
    \mathrm{U\mbox{-}ESR}
    =
    \frac{1}{|D_{\mathrm{test}}|}
    \sum_{x\in D_{\mathrm{test}}}
    \mathbf{I}\!\left(C(\tilde{x})\neq y_{\mathrm{src}}\right).
\end{equation}
Targeted evasion success rate (T-ESR) is the fraction classified as the
selected target class:
\begin{equation}
    \mathrm{T\mbox{-}ESR}
    =
    \frac{1}{|D_{\mathrm{test}}|}
    \sum_{x\in D_{\mathrm{test}}}
    \mathbf{I}\!\left(C(\tilde{x})=y_{\mathrm{tgt}}\right).
\end{equation}
Here $D_{\mathrm{test}}$ is the evaluation set, $\tilde{x}$ is the morphed flow, and $C$ is the evaluated censor classifier. 
For early-stage evaluation, we report these metrics at observation length $k$ by applying $C$ to the morphed early-stage window $\tilde{x}_{1:k}$. 

\noindent\textbf{Traffic validity.}
Validity rate (VR) is the fraction of morphed flows that satisfy protocol and action-space constraints. 
A valid morphed flow preserves packet direction, packet order, and packet count; uses only additive padding and
non-negative delay; respects MTU, bandwidth, and latency budgets; and
introduces no transport-level inconsistencies. 
For end-to-end validation, we further replay generated packet traces with \texttt{tcpreplay}\footnote{\url{https://tcpreplay.appneta.com/}}, inspect them using Wireshark~\cite{beale2006wireshark-capsinfo-wireshark} expert analysis, and check application-layer fingerprints when available. Unless otherwise stated, invalid morphed flows are counted as evasion failures when computing T-ESR and U-ESR.

\noindent\textbf{Operational overhead.}
Bandwidth overhead measures the relative increase in transmitted bytes due to padding, and latency overhead measures the relative increase in flow duration due to added delay:
\begin{equation}
\label{eq:overhead}
    \delta_{\mathrm{bw}}
    =
    \frac{\sum_{t=1}^{T}(\tilde{s}_t-s_t)}
         {\sum_{t=1}^{T}s_t}
    \times 100\%,
    \quad
    \delta_{\mathrm{lat}}
    =
    \frac{\sum_{t=1}^{T}(\tilde{\tau}_t-\tau_t)}
         {\sum_{t=1}^{T}\tau_t}
    \times 100\%.
\end{equation}
Unless otherwise stated, attacks are constrained by maximum bandwidth and latency overheads of 30\% and 50\%, respectively.

\noindent\textbf{Generation latency.}
We report the computational cost of generating perturbations separately from traffic latency overhead. 
For {$\mathcal{I}^{\text{oracle}}$}, we report flow-level generation latency, i.e., the time required to generate actions for a 30-packet window. 
For {$\mathcal{I}^{\text{causal}}$}, we additionally report per-packet decision latency, i.e., the time required to commit one packet action {using only the observed packet history}. 
All latencies are measured on a single NVIDIA RTX 4090 GPU with batch size 1 and averaged over 1{,}000 inferences.

\noindent\textbf{Queueing and load metrics.} 
For the load-stress experiment, we report the effective traffic load, peak queue occupancy, packet-drop rate, and packet-reordering rate. 
The effective load is the maximum directional utilization after morphing, normalized by the per-direction bottleneck capacity $C$.
Peak queue occupancy is reported as the aggregate backlog across the two directional queues, in both packets and buffered bytes.

\section{Experimental Results}
\label{sec:results}

This section presents the experimental evaluation of \sol under the threat model defined in \S\ref{sec:threat_model}. 
We organize the results around the research questions in \S\ref{sec:evaluation}. 
{
Our evaluation follows the two complementary interfaces defined in \S\ref{sec:execution_modes}.
We first use $\mathcal{I}^{\text{oracle}}$ to establish the main controlled upper-bound characterization of early-stage morphability through intra-network evasion (\S\ref{subsec:intra_domain}).
We then evaluate $\mathcal{I}^{\text{causal}}$ (\S\ref{sec:causal_eval}) to examine how \sol performs under history-only, online packet-by-packet execution.
}
{
The subsequent robustness analyses return to $\mathcal{I}^{\text{oracle}}$ so that changes in defender adaptation, network environment, and surrogate architecture can be studied under the same controlled observation setting.
These experiments evaluate adaptive defenses (\S\ref{sec:adaptive_censor}), zero-shot cross-environment generalization (\S\ref{subsec:cross_domain}), and surrogate transferability (\S\ref{subsec:transferability}).
Finally, \S\ref{sec:validity} complements these classifier-level evaluations with end-to-end validity and implementation measurements in the network testbed.
}
Unless otherwise stated, reported numbers are averaged over four independent runs with different random seeds for classifier training and morpher initialization.

\noindent\textbf{Classifier calibration.}
All classifiers are calibrated for early-stage classification within a maximum horizon of $T=30$ packets. 
A sensitivity analysis (Appendix~\ref{sec:appendix_calibration_exp1}) shows that clean classification performance saturates around $T=30$ across all four architectures, beyond which computational cost increases without meaningful accuracy gains. 
We therefore use $T=30$ as the maximum early-stage window.
On unperturbed traffic, all four classifiers achieve Macro F1 above 92\% (Table~\ref{tab:censor_baseline} in Appendix~\ref{sec:appendix_calibration_exp1}), confirming that early-flow metadata provides a strong basis for traffic fingerprinting and that evasion is non-trivial. UniNet achieves the highest accuracy (96.50\%) and the lowest cross-dataset variance ($\sigma=0.70$), making it the strongest evaluated classifier. We therefore use UniNet as the primary local surrogate in several experiments and explicitly evaluate cross-surrogate and cross-classifier transferability in \S\ref{subsec:transferability}.

\subsection{Exp-1: {Full-Window Oracle Intra-Network Evasion (RQ1)}}
\label{subsec:intra_domain}

This experiment evaluates the core morphing capability of \sol under the {full-window oracle interface ($\mathcal{I}^{\text{oracle}}$)}.
The goal is to test whether \sol can generate targeted adversarial traffic within a fixed early-stage window while preserving traffic validity and bounded overhead.
{$\mathcal{I}^{\text{oracle}}$ provides a controlled upper-bound characterization of early-stage morphability and is not interpreted as an online evasion result.}
We evaluate \sol in an intra-network setting, where the censor and the evader operate in the same network environment but use disjoint training data.
This isolates evasion capability from cross-environment distribution shift.
Table~\ref{tab:multidomain_full} reports evasion effectiveness, traffic validity, overhead, and generation latency.
Generation latency denotes the time required to generate perturbations for one 30-packet window under {$\mathcal{I}^{\text{oracle}}$}. 
Detailed per-censor results are provided in Appendix~\ref{sec:appendix_exp2}.

\subsubsection{Analysis across Network Environments}

We consider three environments with distinct traffic characteristics: \textit{VideoPF} dataset captures high-throughput adaptive streaming,
\textit{UTMobileNet} dataset captures high-jitter mobile traffic, and \textit{VANT} dataset captures VPN tunnel traffic. To assess cross-architecture transferability, surrogate-based attacks are optimized against a fixed UniNet surrogate and evaluated against four independently trained target censors. 
Query-based baselines use the same local classifier when applicable. We report mean performance across target censors. 
UniNet is selected as the primary surrogate because it captures both fine-grained packet dynamics and coarse flow-level structure, which provides strong cross-architecture transferability~\cite{UniNet-2025,2017-Asiaccs-AML-model}.
The impact of surrogate choice is examined in \S\ref{subsec:transferability}.

\begin{table*}[t]
\centering
\caption{\textbf{Comprehensive analysis of evasion performance.}
We report evasion effectiveness as mean$\pm$std across four target censors for each network environment.
Traffic validity and overhead are averaged across all three environments and four target censors.
All ESR, validity, and overhead values are percentages. Improvements over the strongest baseline are shown in \textcolor{cyan!50!blue}{blue}.}
\label{tab:multidomain_full}
\resizebox{0.99\textwidth}{!}{%
\renewcommand{\arraystretch}{1.12}
\setlength{\tabcolsep}{4.6pt}
\begin{tabular}{l|cc|cc|cc|c|cc|c}
\toprule
\multirow{3}{*}{\textbf{Method}} 
& \multicolumn{6}{c|}{\textbf{Avg. Evasion Success Rate across 4 Censors, mean$\pm$std (\%)}} 
& \multirow{3}{*}{\textbf{Valid.}} 
& \multicolumn{2}{c|}{\textbf{Overhead}} 
& \multirow{3}{*}{\textbf{Gen. Latency}} \\
\cmidrule(lr){2-7} \cmidrule(lr){9-10}
& \multicolumn{2}{c|}{\textbf{VideoPF}} 
& \multicolumn{2}{c|}{\textbf{UTMobileNet}} 
& \multicolumn{2}{c|}{\textbf{VANT}} 
& 
& \textbf{$\delta_{\text{bw}}$} 
& \textbf{$\delta_{\text{lat}}$} 
& \\
& \textbf{U-ESR} & \textbf{T-ESR} 
& \textbf{U-ESR} & \textbf{T-ESR} 
& \textbf{U-ESR} & \textbf{T-ESR} 
& \scriptsize{(\%)} 
& \scriptsize{(\%)} & \scriptsize{(\%)} 
& \scriptsize{(ms)} \\
\midrule

RandP 
& \msp{14.5}{4.1}
& \msp{4.5}{1.3}
& \msp{12.6}{2.7}
& \msp{3.2}{0.3}
& \msp{11.0}{1.0}
& \msp{3.8}{0.6}
& 100.0
& \msp{26.6}{1.8}
& \msp{24.2}{4.4}
& -- \\

BAP~\cite{BAP-2021-Usenix} 
& \msp{43.8}{13.6}
& \msp{28.6}{12.1}
& \msp{35.9}{5.7}
& \msp{22.2}{5.0}
& \msp{31.8}{2.5}
& \msp{18.5}{1.9}
& \msp{52.7}{5.2}
& \textbf{\msp{21.4}{4.0}}
& \textbf{\msp{16.1}{3.3}}
& \msp{43.1}{5.8} \\

Amoeba~\cite{Amoeba-2023-CoNext} 
& \underline{\msp{84.6}{3.3}}
& \underline{\msp{73.7}{3.8}}
& \underline{\msp{76.6}{3.8}}
& \underline{\msp{64.0}{4.4}}
& \underline{\msp{74.4}{2.9}}
& \underline{\msp{61.7}{2.6}}
& \msp{78.5}{3.3}
& \msp{28.2}{1.2}
& \msp{43.9}{4.3}
& \msp{117.9}{18.6} \\

\textbf{PANTS-BB}~\cite{PANTS-2025-Usenix} 
& \msp{77.7}{6.0}
& \msp{66.8}{5.3}
& \msp{71.5}{6.9}
& \msp{58.2}{9.7}
& \msp{73.4}{4.1}
& \msp{61.6}{5.8}
& \textbf{100.0}
& \msp{25.2}{2.7}
& \msp{32.9}{5.0}
& \msp{57.3}{7.4} \\

\rowcolor{gray!10}
\textbf{\sol}
& \mspg{92.8}{1.7}{+9.7}
& \mspg{85.7}{2.7}{+16.4}
& \mspg{89.7}{1.3}{+17.1}
& \mspg{81.8}{1.5}{+27.7}
& \mspg{90.1}{1.3}{+21.1}
& \mspg{82.0}{1.6}{+32.9}
& \textbf{100.0}
& \msp{26.1}{2.1}
& \msp{38.2}{5.9}
& \textbf{\msp{1.13}{0.09}} \\

\bottomrule
\end{tabular}%
}
\end{table*}

Across all environments, Table~\ref{tab:multidomain_full} shows that targeted impersonation is harder than untargeted confusion when perturbations are transferred from a local surrogate to unseen censors.
Perturbations learned on the surrogate often suffice to move traffic away from its source class, but steering it toward a specific allowed target requires finer control over class-specific decision boundaries. 
While baselines exhibit a wider gap between U-ESR and T-ESR, \sol maintains a consistent gap below 8\% and T-ESR above 80\% across all datasets. 
{On VANT, \sol achieves 82.0\% T-ESR, exceeding the strongest baseline, Amoeba (61.7\%), by 20.3\%.}
This improvement is attributed to the target-conditioned alignment mechanism (Eq.~\ref{eq:attention}), which conditions perturbations on the target class structure instead of relying on generic noise injection.

UTMobileNet serves as a stress test because its high mobile-network jitter weakens precise timing-based perturbations. The T-ESR of solver-based PANTS-BB drops to 58.2\%, reflecting the difficulty of finding robust per-flow solutions under stochastic timing variation.
\sol achieves {81.8\% T-ESR}, indicating that its learned morphing strategy remains effective under noisy network conditions. 
Heuristic and mutation-based approaches suffer either lower targeted evasion or reduced validity, meaning that some successful classifier evasions do not correspond to valid packet sequences. 
PANTS-BB guarantees validity, but its per-flow solver latency remains high. In contrast, \sol amortizes constraint handling during training and reduces morphing to a single forward pass, while the constraint-projection layer enforces valid packet-level actions.

\subsubsection{Per-Censor Breakdown}
\label{subsubsec:videopf_case_study}

\begin{figure}[htbp]
    \centering
    \captionsetup[subfigure]{skip=2pt}
    \begin{subfigure}[b]{0.24\linewidth}
        \centering
        \includegraphics[width=\linewidth]{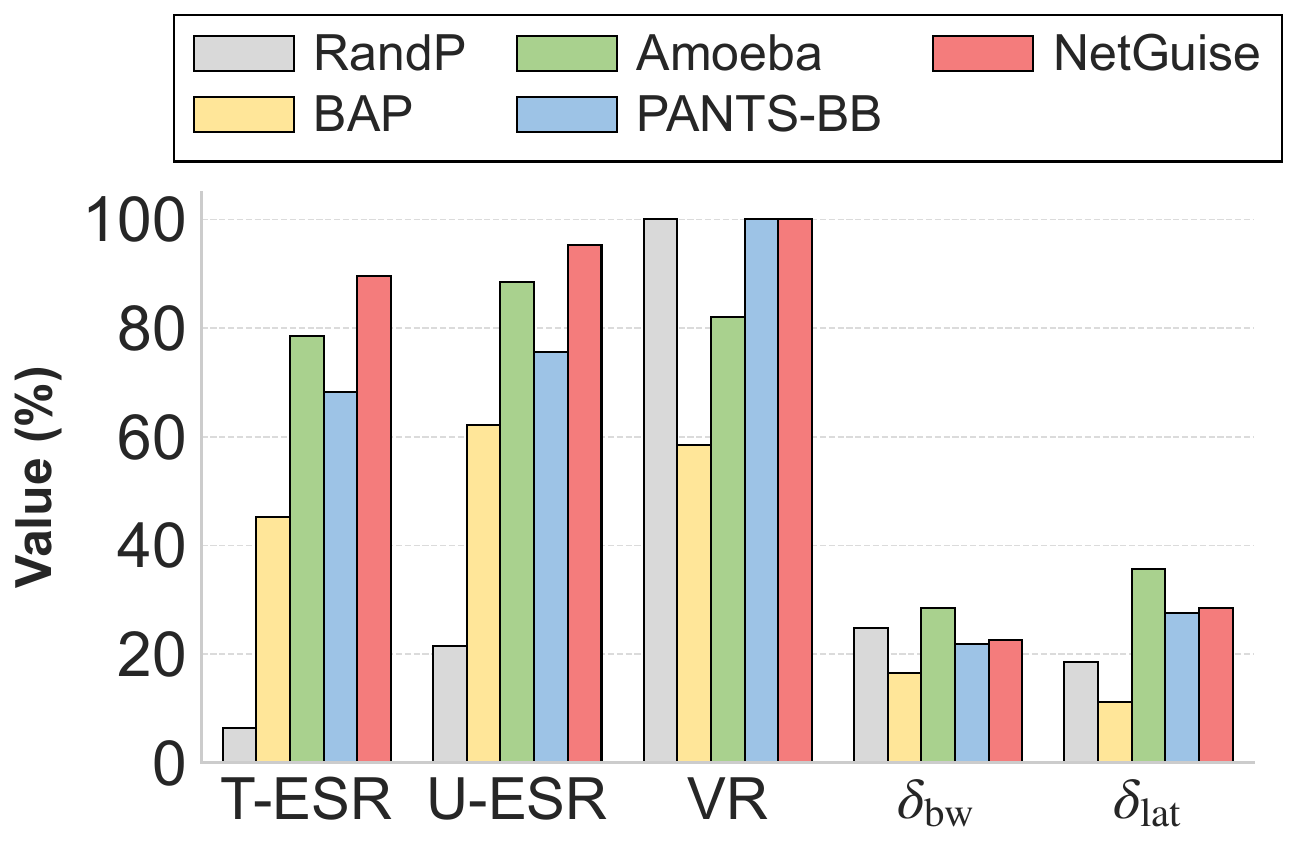}
        \caption{Grad-BP}
        \label{fig:res_gradbp}
    \end{subfigure}
    \hfill
    \begin{subfigure}[b]{0.24\linewidth}
        \centering
        \includegraphics[width=\linewidth]{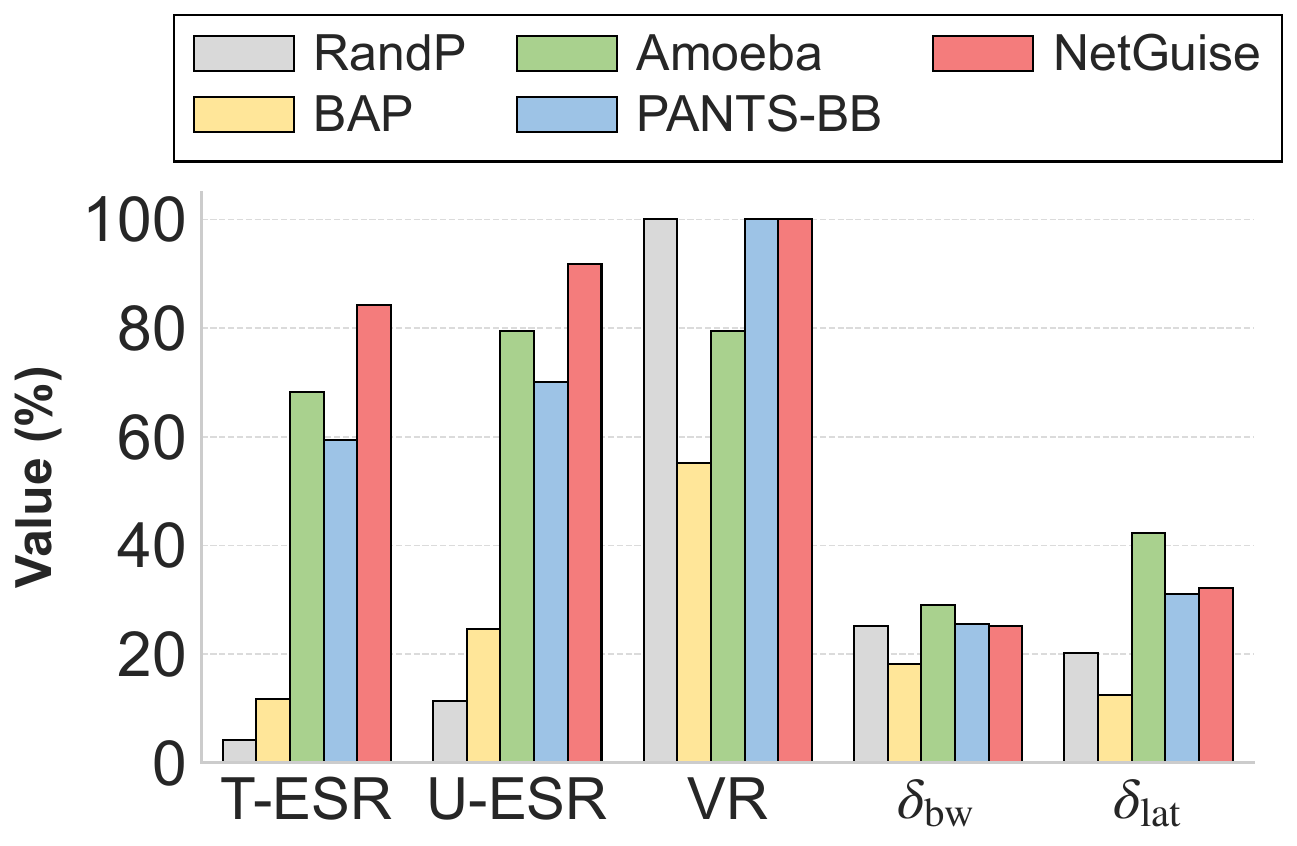}
        \caption{GGFAST}
        \label{fig:res_ggfast}
    \end{subfigure}
    \hfill
    \begin{subfigure}[b]{0.24\linewidth}
        \centering
        \includegraphics[width=\linewidth]{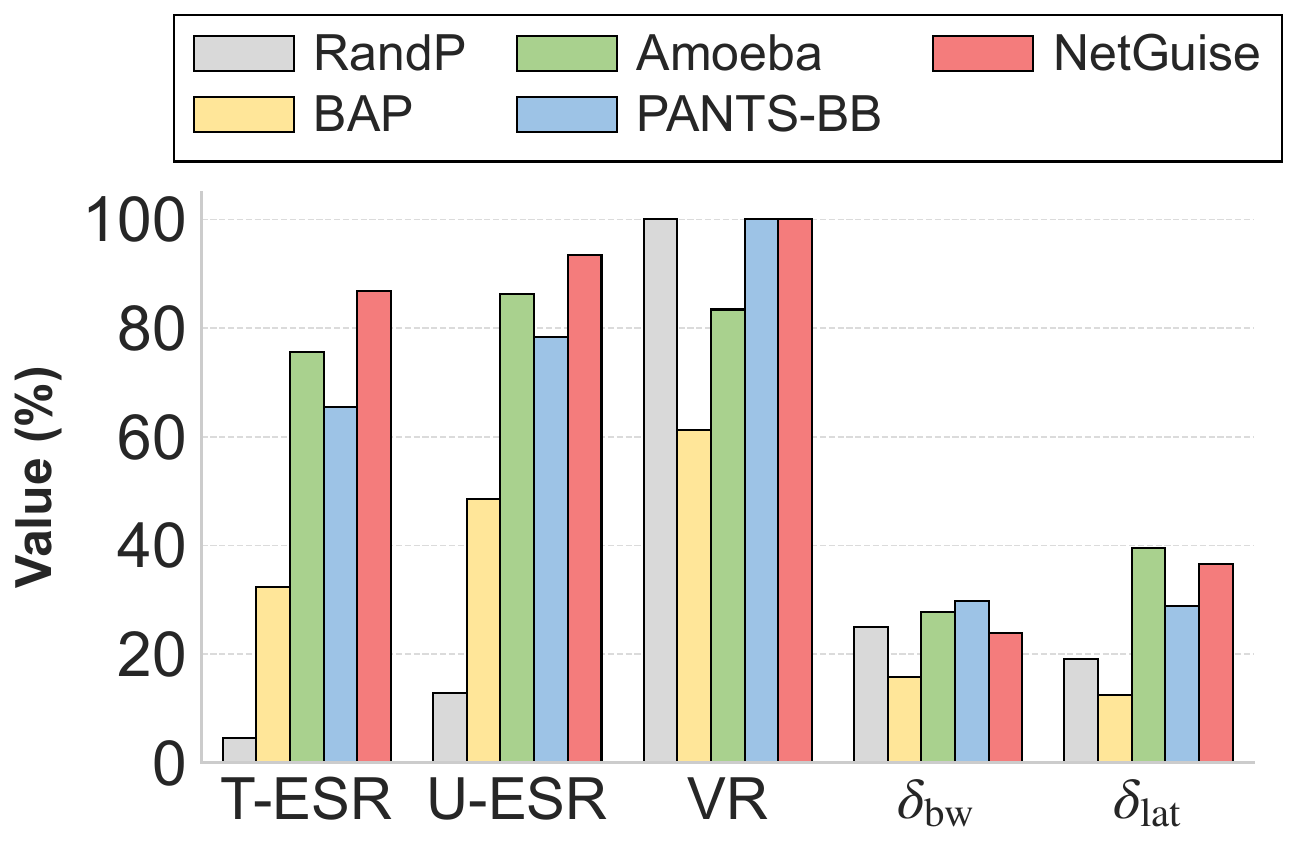}
        \caption{FastFlow}
        \label{fig:res_fastflow}
    \end{subfigure}
    \hfill
    \begin{subfigure}[b]{0.24\linewidth}
        \centering
        \includegraphics[width=\linewidth]{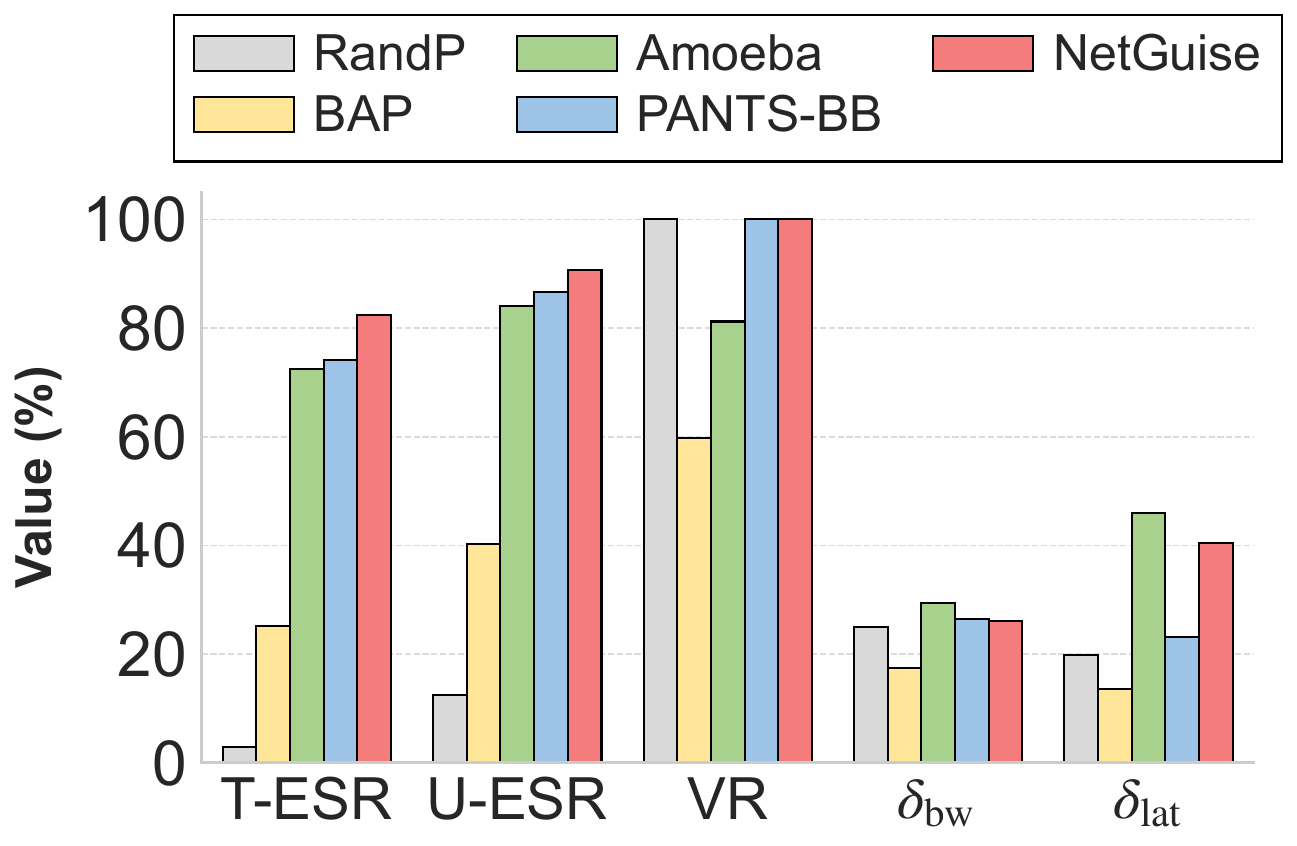}
        \caption{UniNet}
        \label{fig:res_uninet}
    \end{subfigure}

    \caption{{Per-censor full-window oracle evasion on VideoPF-2025.} Metrics include targeted/untargeted evasion success rate
    (T/U-ESR), validity rate (VR), bandwidth overhead ($\delta_{\mathrm{bw}}$), and latency overhead ($\delta_{\mathrm{lat}}$).}
    \label{fig:main_results_1x4}
\end{figure}

We use VideoPF-2025 for the per-censor breakdown because its high-throughput streaming workload imposes a demanding bandwidth-overhead
trade-off: excessive padding quickly increases total transmitted bytes.
Fig.~\ref{fig:main_results_1x4} reports results for each target censor.
Results on additional datasets are provided in Appendix~\ref{sec:appendix_exp2}.

BAP attains T-ESR of at most 45.2\% against Grad-BP with validity below 60\%, confirming that feature-level or weakly constrained perturbations lack the precision required for targeted packet-level impersonation.
Amoeba achieves high U-ESR against some censors, but incurs higher overhead and lower validity than \sol. PANTS-BB performs strongly when the target censor is close to the surrogate, but degrades against heterogeneous architectures, indicating sensitivity to surrogate-specific decision geometry. 
In contrast, \sol maintains T-ESR above 82\% across all four censors, while achieving 100\% validity and the smallest per-censor variance. 
This indicates that \sol learns transferable target-conditioned morphing strategies rather than overfitting to a single surrogate classifier.

\subsubsection{Per-Pair Morphability Analysis}
\label{subsubsec:pair_structure}

The aggregate results show that \sol achieves strong targeted evasion across datasets and censors. 
We next analyze \emph{why} some source--target directions are easier to morph than others, and how \sol uses the additive shaping budget. 
This analysis is important for robustness evaluation: evasion success is not determined only by classifier weakness or attack strength, but also by the structural compatibility between the source and target traffic.

\begin{table}[htbp]
\centering
\caption{Source--target morphability, targeted evasion, and overhead on VANT dataset. }
\label{tab:pair_morphability_evasion}
\resizebox{\textwidth}{!}{%
\begin{tabular}{llcrrr|ccc|cccc}
\toprule
\multicolumn{6}{c|}{\textbf{Source--Target Structure}} &
\multicolumn{3}{c|}{\textbf{T-ESR (\%)}} &
\multicolumn{4}{c}{\textbf{Overhead (\%)}} \\
\cmidrule(lr){1-6}
\cmidrule(lr){7-9}
\cmidrule(lr){10-13}
\textbf{Pair} &
\textbf{Direction} &
\textbf{Regime} &
$\Delta_s$ &
$\Delta_\tau$ &
\textbf{$ C_d$} &
\textbf{\sol} &
\textbf{Best Base.} &
\textbf{Gain} &
$\delta_{\mathrm{bw}}^{\sol}$ &
$\delta_{\mathrm{bw}}^{\mathrm{base}}$ &
$\delta_{\mathrm{lat}}^{\sol}$ &
$\delta_{\mathrm{lat}}^{\mathrm{base}}$ \\
\midrule
A & C2$\to$Streaming
& Size
& +467 & $-17.3$ & 0.68
& \textbf{85.8} & 65.4 (PBB) & +20.4
& 28.5 & \textbf{28.2} & \textbf{30.2} & 35.8 \\

B & C2$\to$Chat
& Timing
& +16 & +43.3 & 0.38
& \textbf{80.2} & 58.8 (PBB) & +21.4
& \textbf{12.8} & 15.4 & 48.5 & \textbf{38.2} \\

C & VoIP$\to$Chat
& Timing
& +38 & +77.2 & 0.35
& \textbf{78.5} & 56.5 (Amo) & +22.0
& \textbf{14.5} & 28.5 & \textbf{45.2} & 48.2 \\

D & VoIP$\to$Streaming
& Size
& +489 & +16.6 & 0.62
& \textbf{85.1} & 65.5 (PBB) & +19.6
& 27.1 & \textbf{26.5} & \textbf{28.4} & 32.1 \\
\bottomrule
\end{tabular}%
}
\end{table}

We focus on four censorship-evasion pairs in VANT dataset: two blocked-like source classes, C2 and VoIP, crossed with two allowed target classes, Streaming and Chat. 
C2 includes tunnel, proxy, and covert-channel traffic, while VoIP is blocked or restricted in multiple jurisdictions. 
Streaming and Chat are selected as allowed targets because blocking them would cause substantial collateral damage. 
Under additive-only morphing, source--target morphability is governed by three factors: 
(i) the packet-size gap $\Delta_s=\bar{s}_{\mathrm{tgt}}-\bar{s}_{\mathrm{src}}$, where positive values favor padding; 
(ii) the timing gap $\Delta_\tau=\bar{\tau}_{\mathrm{tgt}}-\bar{\tau}_{\mathrm{src}}$, where positive values require delay-based reshaping; and 
(iii) direction-sequence compatibility ($ C_d$), which measures the fraction of matching packet directions. 
Direction compatibility is important because packet directions are immutable under our action space.

Table~\ref{tab:pair_morphability_evasion} shows that the four pairs fall into two morphing regimes. 
Pairs A and D are \emph{size-dominated}: the target class has substantially larger packets than the source class, with $\Delta_s>460$ bytes. 
In these directions, additive padding directly moves the source packet-size distribution toward the target, and the higher Dir-Match values reduce immutable direction-level mismatch. 
As a result, \sol achieves the highest T-ESR on these pairs. 
Pairs B and C are \emph{timing-dominated}: source and target packet sizes are already close, but the target class has much larger inter-arrival times and lower direction compatibility. 
These directions require more delay-based reshaping and leave less room for size-based alignment. 
In these harder cases, \sol still achieves 80.2\% and 78.5\% T-ESR, respectively. 
Across all four directions, \sol improves targeted evasion over the strongest baseline by 19.6--22.0\%.
The overhead columns further show how the morphing strategy changes with the source--target regime. 
For size-dominated Streaming targets, \sol uses higher bandwidth overhead because padding is the main mechanism for closing the packet-size gap, while latency overhead remains moderate. 
For timing-dominated Chat targets, \sol uses much lower bandwidth overhead but relies more on delay, reflecting the larger timing gap. 
This coupling between structural compatibility, evasion success, and overhead shows that early-stage robustness should be evaluated as a morphability problem, not only as an aggregate classifier-accuracy problem. 
Detailed structural statistics and full per-pair U-ESR/overhead results are provided in Appendix~\ref{app:pair_structure_full}. The corresponding packet-size and inter-arrival-time distributions are shown in Appendix~\ref{app:pairdist}, where size-dominated pairs exhibit stronger overlap after padding and timing-dominated pairs retain more residual temporal structure.

\subsection{{Exp-2: History-Only Causal Evaluation (RQ2)}}
\label{sec:causal_eval}

Having established the controlled upper-bound characterization under $\mathcal{I}^{\text{oracle}}$, we next evaluate how much of this morphability remains under the history-only causal interface ($\mathcal{I}^{\text{causal}}$).
Under $\mathcal{I}^{\text{causal}}$, the morpher observes only the packet history available so far and must commit each action before later packets are observed.

\begin{figure}[htbp]
    \centering
    \includegraphics[width=0.6\linewidth]{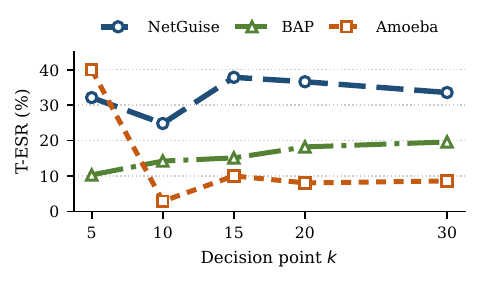}
    \caption{Targeted evasion under the history-only causal interface on VANT across decision points.}
    \Description{Line chart comparing the targeted evasion success rates of NetGuise, BAP, and Amoeba across decision points from 5 to 30 packets.}
    \label{fig:online_tesr}
\end{figure}

We compare \sol with BAP and an online-adapted version of Amoeba.
BAP provides a query-free baseline: its blind perturbations are learned offline and do not depend on future packets or runtime classifier queries.
We restrict BAP to the same padding and non-negative-delay action space and overhead budgets as \sol.
Amoeba was not originally designed for a strict history-only, commit-once interface; we therefore adapt it so that, at packet step $t$, it may modify only the current packet using the observed history and remaining budgets, while retaining its per-packet classifier queries.
\sol operates under the same commit-once execution constraints, but requires no classifier queries at inference time.

For all three methods, the target class is fixed before attack execution, and actions preserve packet direction, packet order, and packet count.
We evaluate decision points $k\in\{5,10,15,20,30\}$ under bandwidth and latency budgets of 30\% and 50\%, respectively.
Future packet positions are masked for \sol as defined in \S\ref{sec:execution_modes}, and only the action for the current packet is committed at each step. Once committed, an action cannot be revised.

Fig.~\ref{fig:online_tesr} shows targeted evasion across decision points.
{\sol maintains relatively stable targeted steering, achieving 24.82--37.86\% T-ESR across the evaluated prefixes. 
BAP exhibits a gradual increase from 11.3\% at $k=5$ to 18.6\% at $k=30$, but remains below \sol at every decision point. 
Averaged across $k$, BAP achieves 15.48\% T-ESR compared with 33.20\% for \sol.}

{This comparison is particularly informative because BAP and \sol operate under the same history-only, query-free information setting.
Their 17.72\% difference in mean T-ESR therefore reflects differences in attack design rather than additional runtime classifier access. 
In particular, BAP applies input-independent blind perturbations, whereas \sol conditions its actions on the observed history and target structure.}
The adapted Amoeba baseline behaves differently. It achieves 40.00\% T-ESR at $k=5$ but drops to 2.86\% at $k=10$ and remains below 10\% for $k\geq20$. 
Despite retaining per-packet classifier feedback, its mean T-ESR is only 13.89\%. 
Thus, stronger runtime classifier access alone does not provide stable targeted steering under commit-once causality.

Table~\ref{tab:online_summary} summarizes the corresponding effectiveness, overhead, and runtime.  
Under the same query-free setting, \sol improves mean T-ESR over BAP by 17.72\% and mean U-ESR by 24.98\%. 
\sol also reduces bandwidth overhead from 19.80\% to 3.02\% and generation latency from 37.42 to 1.16 ms/flow, a 32.3$\times$ speedup. 
BAP, however, incurs substantially lower latency overhead, exposing a trade-off between delay overhead and targeted evasion effectiveness. 
Amoeba operates under a stronger runtime information assumption because it retains per-packet classifier queries, yet achieves only 13.89\% mean T-ESR and requires 136.17 ms/flow. 
Together, these comparisons show that \sol's causal advantage does not come from additional runtime classifier feedback. 
In particular, the comparison with BAP, where both methods operate under the same history-only and query-free setting, provides a more controlled comparison of their attack designs.

\begin{table}[htbp]
    \centering
    \caption{History-only causal evaluation on VANT.}
    \label{tab:online_summary}

    \resizebox{0.5\linewidth}{!}{%
    \begin{tabular}{lccc}
        \toprule
        \textbf{Metric} &
        \textbf{BAP} &
        \textbf{Amoeba} &
        \textbf{\sol} \\
        \midrule
        Mean T-ESR (\%)      & 15.48 & 13.89 & \textbf{33.20} \\
        Mean U-ESR (\%)      & 31.34 & 44.14 & \textbf{56.32} \\
        BW overhead (\%)     & 19.80 & 13.81 & \textbf{3.02} \\
        Lat. overhead (\%)   & \textbf{26.71} & 49.70 & 49.91 \\
        Gen. ms/flow         & 37.42 & 136.17 & \textbf{1.16} \\
        Runtime queries      & None & Per pkt & None \\
        \bottomrule
    \end{tabular}%
    }

\end{table}

We additionally evaluate $\mathcal{I}^{\text{causal}}$ on UTMobileNet to examine a more difficult traffic environment.
All three methods achieve substantially lower targeted evasion than on VANT, confirming that causal morphing is strongly dataset-dependent.
Averaged across decision points, \sol achieves 5.60\% T-ESR, compared with 4.11\% for BAP and 1.63\% for Amoeba. \sol also achieves the highest mean U-ESR at 41.89\%, compared with 30.22\% for BAP and 34.34\% for Amoeba. 
At shorter decision points ($k\leq15$), \sol provides the strongest targeted evasion, whereas BAP becomes slightly
more competitive at longer prefixes and marginally exceeds \sol at $k=20$ and $k=30$.
These results suggest that the benefit of causal morphing depends on both the traffic environment and the available prefix length. 
On UTMobileNet, high variability in early packet sequences makes history-conditioned targeted steering substantially more difficult, while blind perturbations can accumulate effectiveness as a longer prefix becomes available.
In terms of efficiency, \sol reduces bandwidth overhead from 24.52\% to 13.53\% relative to Amoeba, decision latency from 3.843~ms/packet to 0.018~ms/packet, and generation latency from 114.98~ms/flow to 0.54~ms/flow. 
Detailed UTMobileNet results are reported in Appendix~\ref{app:utmobile-online}.

Overall, the causal evaluation shows that history-only targeted morphing remains challenging and strongly dataset- and prefix-dependent. 
Nevertheless, \sol provides the highest average targeted and untargeted evasion on UTMobileNet while requiring no runtime classifier queries.

\subsection{Exp-3: {Robustness Against Adaptive Defenses (RQ3)}}
\label{sec:adaptive_censor}

{
After evaluating online causal execution under $\mathcal{I}^{\text{causal}}$ in \S\ref{sec:causal_eval}, we return to $\mathcal{I}^{\text{oracle}}$ for the remaining controlled robustness analyses.
Exp-1 and Exp-2 evaluate \sol against censors that remain fixed during attack evaluation.
Here, we instead examine how robustness changes when the defender adapts, while keeping the default no-query inference assumption in \S\ref{sec:threat_model}.
}
The evader and defender observe representative outputs from the previous round during retraining, but \sol still performs query-free inference against the deployed classifier. 
This follows the Kerckhoffs-style principle that an evasion method should remain effective even when the defender observes representative adversarial examples and updates its classifier~\cite{2023-sok-kerckhoff}.

{We evaluate three complementary defense settings. First, we consider adaptive retraining, in which the censor updates its decision boundary using previously observed morphed traffic while keeping its observation horizon and feature representation fixed. We then evaluate two orthogonal defenses that modify the classifier interface itself: extending the observation window beyond \sol's 30-packet morphing horizon, and restricting features that can be directly manipulated by padding or delay. Together, these experiments characterize whether robustness can be improved through decision-boundary adaptation, additional observation, or feature invariance.}

Each round~$r$ proceeds in two phases: i)~\textit{evader phase:} the generator $\mathcal{G}_r$ is trained from scratch using differentiable surrogate model $\mathcal{S}_r$. $\mathcal{G}_r$ is then applied to the source test flows, producing a set of morphed flows $\mathcal{M}_r$. We evaluate $\mathcal{G}_r$ against $\mathcal{C}_r$ and record T-ESR and U-ESR for round~$r$; ii)~\textit{censor phase:} the morphed flows $\mathcal{M}_r$ are labeled with their \emph{true source class} (e.g., a C2 flow morphed to resemble Streaming is labeled C2). A size-matched random subsample $\widetilde{\mathcal{M}}_r \subset \mathcal{M}_r$ (with replacement, $|\widetilde{\mathcal{M}}_r| = |\mathcal{D}_{\text{orig}}|$, class-proportional) is mixed with the original training data to form:
    \begin{equation}
        \mathcal{D}_{r+1} = \mathcal{D}_{\text{orig}} \;\cup\; \widetilde{\mathcal{M}}_r
        \label{eq:adaptive_data}
    \end{equation}
\noindent A new censor classifier $\mathcal{C}_{r+1}$ is trained from scratch on $\mathcal{D}_{r+1}$. 
The total training set size $|\mathcal{D}_{r+1}| = 2|\mathcal{D}_{\text{orig}}|$ is held constant across rounds. Only the morphed flows from the most recent round are included (not accumulated from prior rounds), isolating adaptation effect of past rounds.
Under this setting, both sides observe representative outputs from the previous adaptation round. The key question is whether the generator can continue discovering viable evasion strategies as the coverage of the morphing space of censors expands.

\subsubsection{Adaptive Results Across Rounds}
\begin{figure}[htbp]
    \centering
    \includegraphics[width=0.72\linewidth]{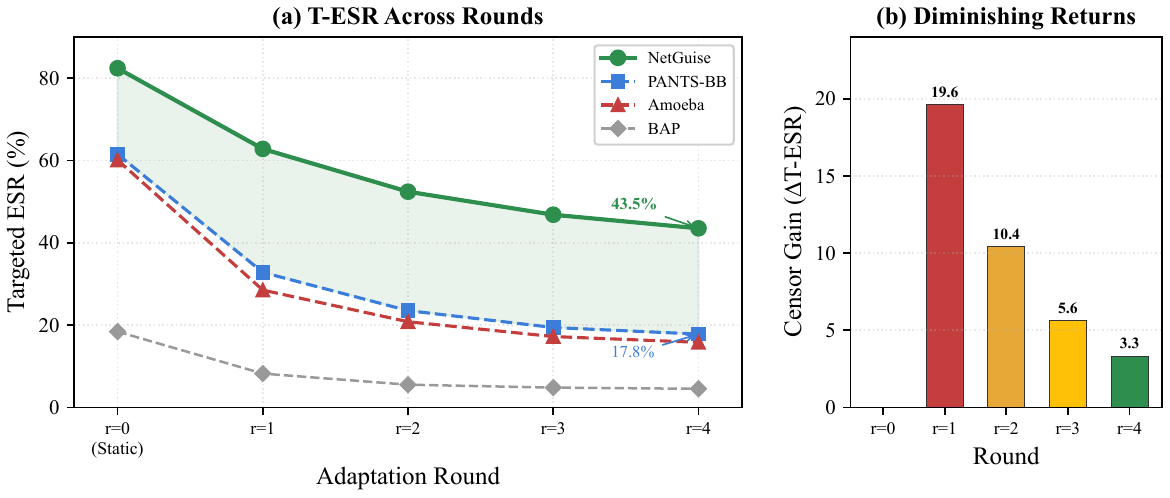}
    \caption{T-ESR across four rounds of iterative censor adaptation.}
    \Description{Two-panel plot showing targeted evasion success rates across adaptation rounds and the diminishing gains obtained by iterative censor retraining.}
    \label{fig:multiround}
\end{figure}

Fig.~\ref{fig:multiround}a tracks T-ESR across all four rounds; Fig.~\ref{fig:multiround}b illustrates the percentage gain in censor performance after each round. The detailed T-ESR and U-ESR results across four adaptation rounds, averaged across all VANT pairs, are shown in Table~\ref{tab:multiround} (Appendix~\S\ref{exp2-5-rounds}).

\noindent\textbf{Single-round analysis ($r{=}0 \to r{=}1$)}
After the first round of adaptation, RandP and BAP are nearly neutralized: BAP's T-ESR drops from 18.4\% to 8.2\%. This result means that low-diversity perturbations are fully learnable.
Amoeba and PANTS-BB initially achieve  T-ESR of $\sim$60\% but suffer significant degradation (31.6\% for Amoeba, 28.7\% for PANTS-BB). 
Moreover, the gap between them narrows after adaptation, suggesting that distinguishing features of each method become less relevant once the censor has seen representative morphed examples.
In contrast, \sol's T-ESR decreases by 19.6\% to 62.8\%, and its U-ESR remains at 72.5\% (vs.\ 48.5\% for the best baseline), meaning nearly three-quarters of morphed flows still evade source identification.
Despite $\mathcal{C}_1$ having been specifically trained on $G_0$'s morphed flows, $G_1$ of \sol, retrained against $\mathcal{C}_1$, discovers new morphing strategies that exploit gaps in the adapted decision boundary of $\mathcal{C}_1$. The T-ESR gap between \sol and baselines widens from $\sim$21\% to $\sim$30\%, indicating that \sol's architecture has greater capacity to find new evasion strategies when the current ones are blocked.

\noindent\textbf{Multi-round trends ($r{=}1 \to r{=}4$).}
Three insights are revealed.
First, \textit{T-ESR decreases monotonically for all methods}, confirming that the censor has a structural advantage in this arms race: each round, the censor observes the generator's latest output and learns to recognize it. BAP collapses to near-random by $r{=}2$ (5.5\%). Amoeba and PANTS-BB converge to $\sim$17\% by $r{=}4$. \sol degrades more slowly, retaining 43.5\% at $r{=}4$, which is over 2.4$\times$ the best-performing baseline.

 Second, the censor's \emph{marginal gain per round diminishes}. Against \sol, the censor classifier recovers T-ESR of 19.6\% at $r{=}1$, but only 10.4\% at $r{=}2$, 5.6\% at $r{=}3$, and 3.3\% at $r{=}4$ (Fig.~\ref{fig:multiround}b). Each round, the classifier eliminates the most exposed morphing patterns, but $\mathcal{G}_r$ exploits the distributional tail. Covering these residual strategies requires the classifier to tighten its boundary aggressively, increasing the risk of false positives on legitimate allowed traffic.
Third, \sol's relative advantage widens over rounds. The T-ESR gap over PANTS-BB grows from 20.9\% ($r{=}0$) to 25.7\% ($r{=}4$). 
By $r{=}4$, both baselines have converged to nearly identical performance ($\sim$17\%), indicating that the censor has effectively covered perturbation spaces of both methods.

\subsubsection{{Defenses Beyond Adaptive Retraining}}
\label{sec:adaptive-defense-beyond-retraining}

{The adaptive-retraining experiment above changes the censor's decision boundary while keeping its observation horizon and feature representation fixed. 
We therefore further examine two orthogonal defenses that change the information available to the censor: i)~extending the observation window beyond the first 30 packets modified by \sol, and ii)~restricting the input representation to features that are partially or fully invariant to \sol's shaping actions. 
For each extended observation window, the censor is retrained and evaluated using the corresponding packet horizon under the same data split and architecture.
Table~\ref{tab:defense-summary} summarizes the main results, with detailed T-ESR and U-ESR results provided in
Appendix~\ref{sec:defense_features}.}

\begin{table}[htbp]
    \centering
    \caption{Robustness under defenses beyond adaptive retraining on VANT.}
    \label{tab:defense-summary}

    \resizebox{0.6\linewidth}{!}{%
    \begin{tabular}{l l cc}
        \toprule
        \textbf{Defense} &
        \textbf{Setting} &
        \textbf{Clean F1 (\%)} &
        \textbf{T-ESR (\%)} \\
        \midrule
        Default early-stage
        & 30 packets
        & 97.2
        & 82.4 \\

        Longer observation
        & 50 packets
        & \textbf{98.8}
        & 43.6 \\

        Longer observation
        & 100 packets
        & 97.7
        & 26.1 \\

        Padding-invariant
        & Dir. + IAT
        & 62.1
        & 42.3 \\

        Direction-only
        & Direction
        & 41.5
        & \textbf{2.74} \\
        \bottomrule
    \end{tabular}%
    }
\end{table}

\noindent\textbf{Longer observation windows.}
{We first allow the defender to observe beyond the 30-packet morphing window, while \sol continues to modify only the first 30 packets and leaves later packets unchanged. 
Extending the observation window from 30 to 50 packets reduces T-ESR from 82.4\% to 43.6\%, while a 100-packet window further reduces it to 26.1\%. Importantly, clean Macro-F1 remains high across all three settings, ranging from 97.2\% to 98.8\%. 
The reduction in evasion shows packets beyond the attack window retain source-specific information that is unavailable to the morpher. 
Thus, extending the observation horizon provides an effective defense when the censor can tolerate a later classification decision and maintain per-flow state over a longer prefix.}

\noindent\textbf{Feature-restricted defenders.}
We next evaluate classifiers that remove features directly affected by \sol's shaping actions. Removing packet size yields a \emph{padding-invariant} representation based on direction and IAT, reducing T-ESR from 82.4\% to 42.3\%. 
A stricter direction-only representation is invariant to both padding and delay under our action space and further reduces T-ESR to 2.74\%. 
This increased robustness, however, comes at a substantial cost in clean classification performance: Macro-F1 decreases from 97.2\% for the original representation to 62.1\% for direction + IAT and 41.5\% for direction only.

These results expose two distinct robustness trade-offs. 
Extending the observation horizon substantially reduces evasion while preserving high clean classification performance, but sacrifices earlier decision making and requires longer per-flow observation. 
In contrast, restricting the feature representation can remove a larger fraction of \sol's attack surface, but also discards discriminative information and substantially reduces clean classification accuracy. 
Together with adaptive retraining, these results show that robustness against early-stage traffic morphing depends jointly on the classifier's decision boundary, observation horizon, and feature representation.

\subsection{Exp-4: Zero-Shot Cross-Environment Generalization (RQ1)}
\label{subsec:cross_domain}

This experiment tests whether the generator learns transferable adversarial features or overfits to environment-specific artifacts.
There is an asymmetry between censors and adversaries: censors are trained on traffic collected from the \emph{target environment}, whereas evasion systems are typically developed using data from more accessible but mismatched settings.
This motivates a challenging yet realistic zero-shot cross-environment evaluation, in which the censor and the morpher are trained on different network environments, and no domain adaptation or fine-tuning is permitted at deployment time.

\begin{figure}[htbp]
    \centering
    \includegraphics[width=0.7\linewidth]{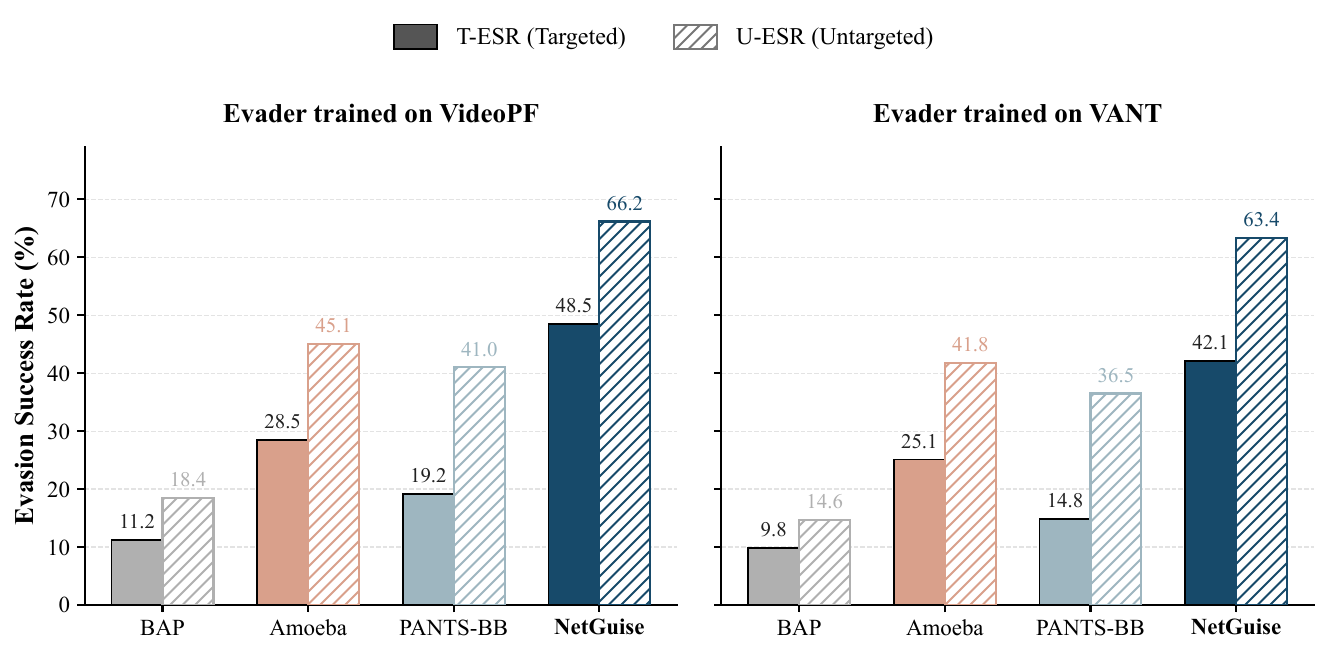}
    \caption{Zero-shot cross-environment evasion performance.}
    \Description{Grouped bar charts comparing targeted and untargeted evasion success rates for BAP, Amoeba, PANTS-BB, and NetGuise when the evader is trained on VideoPF or VANT.}
    \label{fig:zeroshot_scatter}
\end{figure}

In this experiment, the censor is trained on the \textit{UTMobileNet} dataset using the strongest available classifier ({\textit{UniNet}}).
The user, however, is assumed to have \emph{no access} to mobile traffic during training.
Instead, the morpher is trained exclusively on either \textit{VideoPF} or \textit{VANT}, both of which are easier to obtain in practice.
The trained morpher is then deployed directly on mobile traffic without any form of adaptation, constituting a strict zero-shot setting.
Fig.~\ref{fig:zeroshot_scatter} summarizes the resulting T-ESR and U-ESR under this severe environment mismatch.

As shown in Fig.~\ref{fig:zeroshot_scatter}, zero-shot deployment induces a degradation in \emph{targeted impersonation} performance across all methods.
For \sol{}, T-ESR decreases to 48.5\% when trained on VideoPF and to 42.1\% when trained on VANT, highlighting the difficulty of precisely matching mobile-specific timing signatures using non-mobile training data.
This degradation is reasonable: while high-level application structure may transfer across environments, fine-grained packet timing is strongly shaped by environment-specific factors such as radio scheduling and MTU variability.
However, \sol{} retains strong \emph{untargeted evasion} capability, achieving U-ESR of 66.2\% (VideoPF source) and 63.4\% (VANT source).
These results exceed the strongest baseline by more than {45\% relative improvement} in both cases, indicating that \sol{} captures environment-invariant traffic structure to disrupt the censors, even when precise class-level impersonation is imperfect.

\subsection{Exp-5: Surrogate Model Transferability (RQ1)}
\label{subsec:transferability}

Under a strict black-box threat model, evasion systems must be trained against surrogate classifiers whose architecture may differ from the deployed censor.
The experiment here evaluates whether \sol{}’s effectiveness depends on the choice of surrogate architecture, and how surrogate capacity affects cross-architecture generalization.
We train \sol{} using two representative and differentiable surrogates: FastFlow and UniNet.
The resulting morphers are evaluated against four censors across three environments.

\begin{figure}[htbp]
    \centering
    \includegraphics[width=0.9\linewidth]{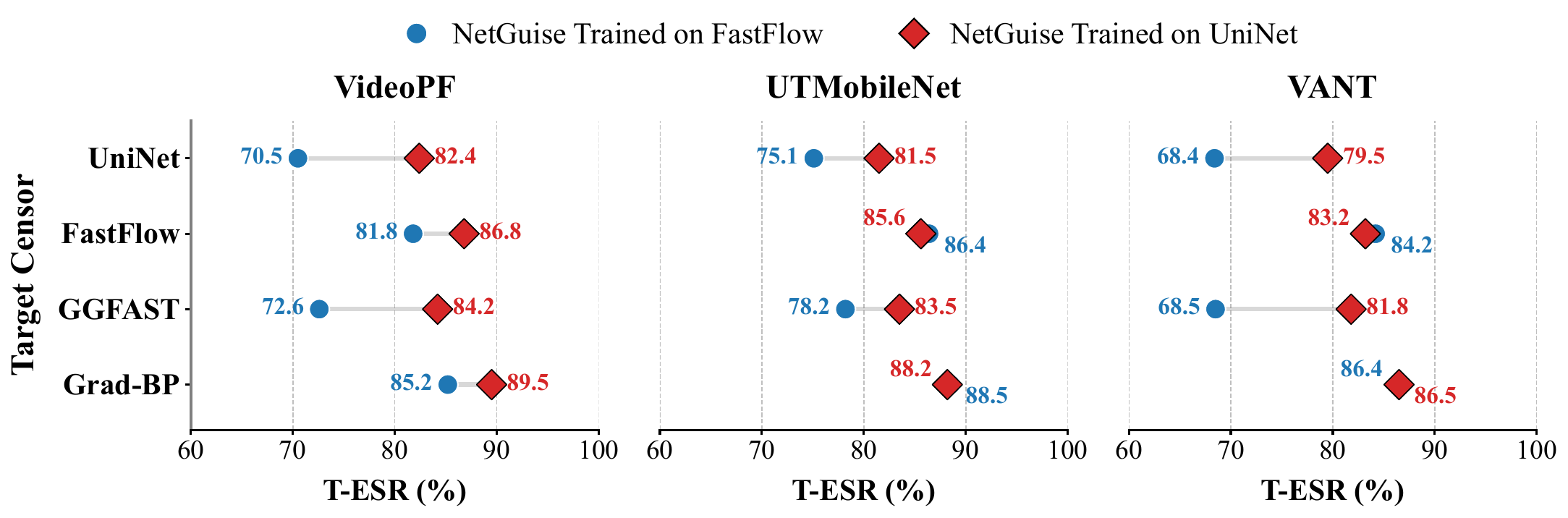}
        \caption{\textbf{Impact of surrogate architecture on \sol{}'s transferability.} 
        Targeted ESR comparison between models trained on FastFlow (Blue Circles) vs. UniNet (Red Diamonds).}
        
        \label{fig:transferability}
\end{figure}

Fig.~\ref{fig:transferability} reports the cross-architecture transferability results.
Across all environments, perturbations trained on the higher-capacity UniNet surrogate consistently achieve stronger and more stable targeted evasion when transferred to diverse black-box censors.
For instance, UniNet-trained morphing attains T-ESR of 84.2\% against GGFAST on VideoPF, compared to 72.6\% when trained on FastFlow, with similar observations across mobile and VPN traffic.
Perturbations optimized on the FastFlow surrogate generalize poorly to more expressive defenders, particularly Transformer-based censors.
This asymmetry indicates that attacks tailored to local temporal dependencies fail to disrupt long-range contextual reasoning.

{\textbf{White-box upper bound.}
When evaluated against the exact surrogate used for training, \sol achieves 89.2--91.4\% T-ESR with FastFlow and 92.7--93.9\% with UniNet across the three datasets (Table~\ref{tab:whitebox_upper} in Appendix~\ref{app:whitebox_upper}). Compared with independently trained same-architecture censors, this yields a transfer gap of 5.8--14.4\%, showing that both intrinsic early-stage morphability and surrogate transfer contribute to the observed evasion performance.}

\subsection{Exp-6: Validity and Live Prototype (RQ4 \& 5)}
\label{sec:validity}

\subsubsection{End-to-End Replay Validation}
Beyond classifier-level evasion metrics, the morphed flows must be packet-level valid and executable on real network stacks.
We evaluate three properties:
protocol-level validity, application-layer fingerprint preservation, and end-to-end execution.

\begin{table}[htbp]
\centering
\renewcommand{\arraystretch}{1.15}

\caption{Combined validity assessment of a \sol-generated morphed flow.
Top: replay statistics (\texttt{tcpreplay}).
Bottom: protocol semantics (\texttt{Wireshark}).}
\label{tab:validity_combined_new}

\resizebox{0.65\linewidth}{!}{%
\begin{tabular}{|l|l|l|}
\hline

\multicolumn{3}{|c|}{\textbf{Physical Replayability (\texttt{tcpreplay})}} \\ \hline
\textbf{Metric} & \textbf{\sol (Morphed)} & \textbf{Original} \\ \hline
Total packets sent & 411 & 411 \\ \hline
Total bytes sent & 69,485 bytes & 66,965 bytes \\ \hline
Successful packets & \textbf{411 (100\%)} & 411 (100\%) \\ \hline
Failed packets & 0 & 0 \\ \hline
Truncated packets & 0 & 0 \\ \hline
Retried (ENOBUFS / EAGAIN) & 0 / 0 & 0 / 0 \\ \hline

\multicolumn{3}{|c|}{\textbf{Protocol Semantics (\texttt{Wireshark})}} \\ \hline
\textbf{Check} & \textbf{\sol (Morphed)} & \textbf{Original} \\ \hline
TCP handshake (SYN/SA/ACK) & \textbf{Verified} (\#1, \#2, \#3) & Verified \\ \hline
Sequence number consistency & No errors & No errors \\ \hline
Expert warnings & \textbf{0} & 0 \\ \hline
TLS session establishment & Valid (ClientHello + ServerHello) & Valid \\ \hline
Application payload & 35,743 bytes & 33,218 bytes \\ \hline

\end{tabular}%
}

\end{table}

\noindent\textbf{Validity check.} 
Table~\ref{tab:validity_combined_new} illustrates the detailed breakdown for a representative flow (VPN tunnel impersonating YouTube streaming). Across the full evaluation set (all morphed flows in all three datasets), we observe zero Wireshark expert warnings and 100\% \texttt{tcpreplay} success with no failed, truncated, or retried packets.
The morphed flow conforms to standard \texttt{pcap} encapsulation, preserves strict timestamp ordering, and contains a valid TCP three-way handshake (\texttt{SYN}, \texttt{SYN-ACK}, \texttt{ACK}) with consistent sequencing and intact encrypted payload delivery.
\texttt{tcpreplay} achieves a \textit{100\% replay success rate} with zero dropped, retried, or truncated packets.

\noindent\textbf{Application-Layer Fingerprint.}
Deployed DPI systems may employ application-layer fingerprinting such as JA3~\cite{ja3} and JA4~\cite{ja4} to identify TLS implementations based on the ClientHello structure (cipher suites, extensions, elliptic curves, and their ordering).
If morphing alters these fingerprints, the resulting traffic could be flagged as anomalous~\cite{ja4}.
We extract JA3 and JA4 hashes from all morphed and corresponding original flows across all three datasets.
All morphed flows preserve identical JA3 and JA4 hashes (100\% match rate for both client and server fingerprints).
The morphing operations in \sol do not alter the TLS ClientHello structure, cipher suite ordering, or extension fields that determine application-layer fingerprints.
{However, fingerprint preservation also introduces a potential detection trade-off. While the packet-level metadata may be shifted toward a target application, the preserved TLS fingerprint may remain characteristic of the source application. A protocol-aware censor could therefore exploit such cross-layer inconsistency between packet metadata and TLS fingerprints as an additional detection signal. Our current evaluation focuses on metadata-based classifiers and does not incorporate such cross-layer consistency checks.}

\noindent\textbf{End-to-End Replay Validation.}
The preceding experiments validate \sol's output. A stronger test is whether
morphed flows remain protocol-valid and classifier-evasive after traversal through a software network testbed with kernel-level packet processing and wire-level transmission.

\begin{figure}[htbp]
    \centering
    \includegraphics[width=0.75\linewidth]{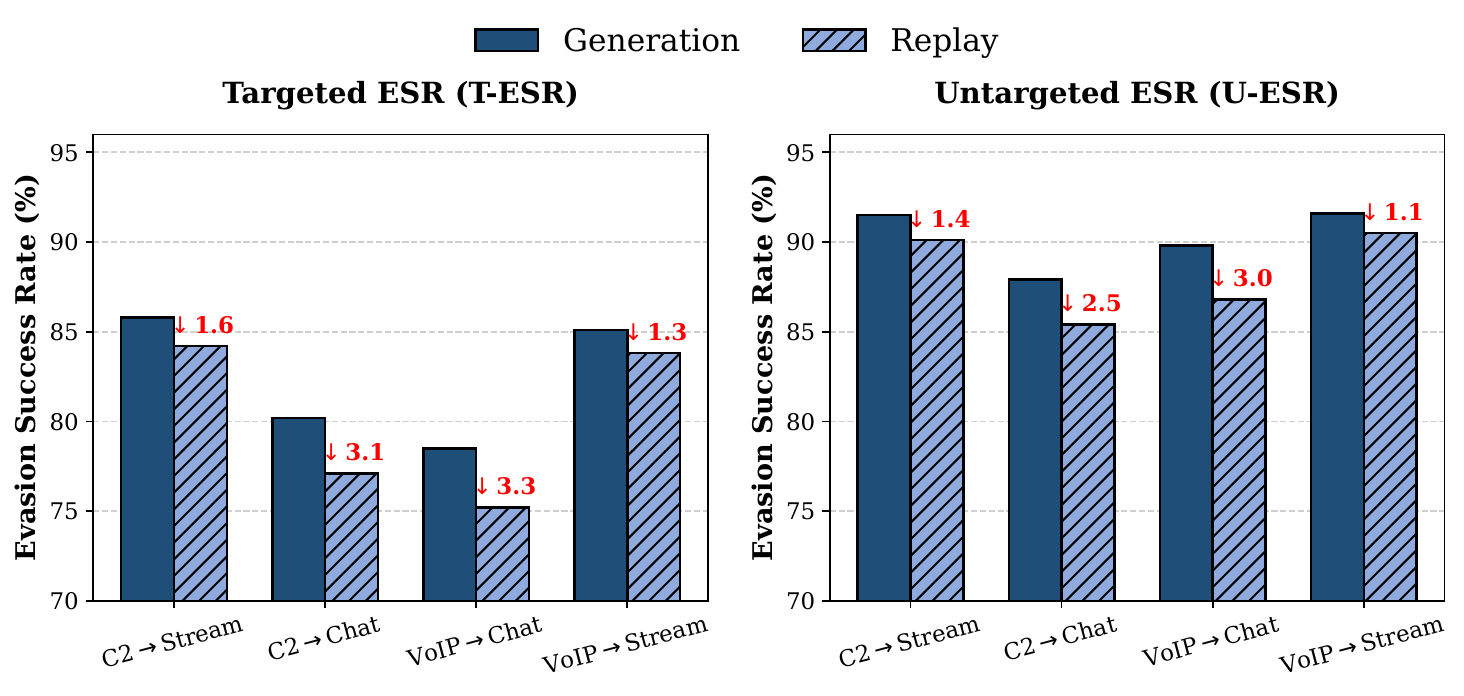}
    \caption{Replay validation on VANT dataset with UniNet as surrogate. Generation denotes direct evaluation, while Replay denotes performance after testbed traversal.}
    \Description{Comparison of targeted and untargeted evasion rates before and after replay through the Mininet testbed for four source--target traffic pairs.}
    \label{fig:replay}
\end{figure}

We construct a controlled network environment using Mininet~\cite{miniNet} and the Ryu SDN controller~\cite{rui2014ryu}. An OpenFlow~1.3 switch~\cite{OpenFlow13} connects a client host, a transparent proxy, and a server host. Packets traversing the switch's client-facing port are captured to represent the vantage point of an on-path censor.
The capture point mirrors the vantage point of a censor: all packets traversing the switch's client-facing port are recorded for classification. This topology allows us to replay \sol-morphed flows through a real TCP/IP stack and observe exactly what an on-path censor would see.
We select 200 flow samples (50 per VANT source$\to$target pair) morphed by \sol, targeting the same four pairs evaluated in \S\ref{subsec:intra_domain} and \S\ref{sec:adaptive_censor}. 
Each flow is replayed through the Mininet testbed via \texttt{tcpreplay}. 
Packets captured at the censor's vantage point are then evaluated against all four censor classifiers. 
As a control, we replay the corresponding \emph{unmorphed} source flows through the same testbed.

\noindent Fig.~\ref{fig:replay} compares evasion performance between morphed flows (evaluated directly) and the same flows after replaying on real network testbed. Replayed flows achieve a T-ESR of 80.1\%, with full protocol validity. 
The per-pair breakdown reveals that size-dominated pairs (A,~D) degrade by only $\sim$1.5\%, while timing-dominated pairs (B,~C) degrade by $\sim$3\%. 
This gap is attributable to kernel-level timing granularity and scheduling jitter ($\sim$2\,ms), which compounds for pairs where evasion relies on precise inter-packet timing. 
 T-ESR of passthrough mode is only 3.2\%, indicating that evasion is attributable to \sol, not to artifacts of the network topology.

\subsubsection{Queueing and Load Stress Test}
\label{sec:queue_stress}

The replay experiment above verifies that traffic generated by \sol remains executable on a real network stack. 
We next examine how the altered packet sizes and release timings of morphed traffic interact with a capacity-constrained network bottleneck under concurrent load.
Padding increases the transmitted byte volume, while non-negative delay changes packet release times and may introduce short-term traffic bursts.
When multiple morphed flows share a bottleneck, these effects can increase queue occupancy and eventually cause packet loss even when each individual flow satisfies its configured shaping budgets.

We extend the same Mininet testbed with a controlled load-stress experiment using VANT.
We configure a bidirectional bottleneck with a per-direction capacity of $C=50$~Mbps.
Each direction uses an HTB rate limiter followed by a FIFO (\texttt{pfifo}) queue with a limit of 256 packets.
Concurrent traffic is replayed at offered-load levels $\rho\in\{25,50,75,100,125\}\%$.
Each run concurrently replays 20 VANT flows using the first 30 packets of each flow, and each condition is repeated three times.
For each direction, traffic load is computed relative to the per-direction bottleneck capacity $C$.
The \emph{offered load} is the maximum directional utilization before morphing, while the \emph{effective load} is the maximum directional utilization after morphing.
Thus, an effective load above 100\% indicates that at least one direction exceeds its configured bottleneck capacity.
\emph{Peak queue occupancy} is sampled every 10~ms and reported as the maximum sampled aggregate backlog across the two directional queues, in both packets and buffered bytes.
\emph{Drop rate} is the fraction of packets dropped at the configured bottleneck queues, and \emph{reordering rate} is measured from the delivered packet order.

\begin{table}[htbp]
    \centering
    \caption{{Queueing and load-stress results on VANT.
Offered and effective loads report the maximum directional utilization, normalized by the per-direction bottleneck capacity $C$.}}
    \label{tab:queue-stress}

    \resizebox{0.55\linewidth}{!}{%
    \begin{tabular}{c c c c c c}
        \toprule
        \textbf{Offered} &
        \textbf{Effective} &
        \multicolumn{2}{c}{\textbf{Peak Queue}} &
        \textbf{Drop} &
        \textbf{Reorder} \\
        \cmidrule(lr){3-4}
        \textbf{Load (\%)} &
        \textbf{Load (\%)} &
        \textbf{Pkts} &
        \textbf{KB} &
        \textbf{(\%)} &
        \textbf{(\%)} \\
        \midrule
        25  & 30.9  & 11  & 12.4  & 0.00 & 0.00 \\
        50  & 61.8  & 27  & 30.1  & 0.00 & 0.00 \\
        75  & 92.9  & 83  & 92.6  & 0.00 & 0.00 \\
        100 & 123.7 & 217 & 240.4 & 1.84 & 0.00 \\
        125 & 154.3 & 356 & 401.7 & 4.26 & 0.00 \\
        \bottomrule
    \end{tabular}%
    }
\end{table}

Table~\ref{tab:queue-stress} shows that queueing remains modest at low and moderate offered loads.
At 25\% and 50\% offered load, peak occupancy is 11 and 27 packets (12.4 and 30.1~KB), respectively, with no packet loss or reordering.
At 75\%, the effective load increases to 92.9\% and the peak queue occupancy grows to 83 packets (92.6~KB), but no packet loss is observed.
This indicates that the configured buffering is sufficient to absorb transient queue buildup under this near-capacity operating condition.

Once the effective traffic demand exceeds the bottleneck capacity, the impact becomes substantially more pronounced.
At 100\% offered load, morphing increases the effective load to 123.7\%, with a peak aggregate queue of 217 packets (240.4~KB) and a 1.84\% drop rate.
At 125\%, the effective load further increases to 154.3\%, while peak queue occupancy reaches 356 packets (401.7~KB) and the drop rate rises to 4.26\%.
No packet reordering is observed at any evaluated load, consistent with \sol's order-preserving shaping constraints.

The testbed remains loss-free through 75\% offered load, including a near-capacity operating point with 92.9\% effective load.
In the evaluated settings, packet loss is observed only after the post-morphing traffic demand exceeds the configured per-direction bottleneck capacity.
These results indicate that finite buffering can absorb transient bursts below capacity, while sustained overload requires additional bandwidth headroom to avoid packet loss.

\section{Discussion}
\label{sec:discussion}

\noindent\textbf{Why early-stage classifiers are vulnerable.}
Our results show that early-stage traffic classifiers can remain vulnerable even when adversarial traffic is restricted to conservative, realizable shaping actions.
{This vulnerability cannot be explained solely by weak classifiers or feature-space artifacts: it appears across multiple censor architectures, network environments, and black-box transfer settings.}
The core issue is structural.
Early-stage classifiers must decide from a short metadata sequence, while the evader can still reshape packet sizes and timings within the permitted tunnel-shaping constraints.
Our per-pair analysis in \S\ref{subsubsec:pair_structure} shows that evasion is highly source--target dependent.
Positive packet-size gaps are especially favorable because additive padding can move source traffic toward the target distribution without shrinking packets.
Direction compatibility further determines how much budget can be spent on size and timing alignment rather than compensating for immutable structural mismatches.
Thus, robustness should not be assessed only by clean accuracy or aggregate evasion rates: specific source--target directions can remain highly morphable even when the classifier appears accurate overall.

\noindent\textbf{Causal execution and query assumptions.}
The gap between $\mathcal{I}^{\text{oracle}}$ and $\mathcal{I}^{\text{causal}}$ shows that access to future packets has a substantial effect on early-stage morphability.
The oracle setting provides a controlled upper-bound characterization, whereas $\mathcal{I}^{\text{causal}}$ requires each action to be committed using only the observed packet history and therefore supports online packet-by-packet execution.
The lower causal performance, particularly on UTMobileNet, shows that targeted morphing becomes substantially more difficult when future packet information is removed.
Under this stricter setting, the comparison with BAP is particularly informative because both methods operate under the same history-only and query-free information budget.
\sol achieves higher targeted evasion than BAP on both VANT and UTMobileNet, indicating that its causal advantage is not due to additional runtime classifier access.
Online-adapted Amoeba provides a complementary comparison: it retains per-packet classifier queries but achieves less stable targeted steering despite this stronger runtime-information assumption.
This distinction matters operationally because query-driven mutation introduces additional latency and assumes access to a classifier interface at inference time.

\noindent\textbf{Implications for adaptive censors.}
Adaptive retraining reduces evasion but does not eliminate it.
Across four rounds, \sol's targeted evasion decreases from 82.4\% to 43.5\%, showing that the censor learns many exposed morphing patterns.
However, the censor's marginal gains diminish over rounds, and \sol retains higher targeted evasion than the evaluated baselines.
This suggests that adversarial retraining alone is an incomplete defense against targeted early-stage morphing.
Our defense experiments further show that retraining is only one axis of robustness.
Extending the observation horizon substantially reduces evasion while preserving high clean classification performance, whereas restricting directly manipulable features can provide stronger resistance at a substantial cost in clean accuracy.
These results expose a broader trade-off among early decision making, feature informativeness, and resistance to traffic morphing.
For censorship systems, the defense problem therefore involves several trade-offs.
Tightening the decision boundary around morphed traffic can reduce evasion, but may also increase false positives on legitimate allowed applications that occupy nearby metadata regions.
This risk is especially pronounced for size-dominated pairs, where padding creates direct distributional overlap with the target class.
More robust defenses may require features less sensitive to additive padding and delay, such as higher-order temporal structure, direction-sequence constraints, protocol-level behavioral fingerprints, or active probing.
These defenses may raise attack cost, but they also introduce deployment costs in latency, complexity, and collateral damage.

\noindent\textbf{Limitations.}
We instantiate and evaluate \sol primarily in the use case of censorship evasion, which represents one application of early-stage metadata-based traffic classification.
Our threat model focuses on censorship decisions based on packet metadata.
It does not model orthogonal mechanisms for detecting the tunnel itself, such as endpoint IP addresses, SNI, TLS fingerprints, or active probing.
These signals could provide additional detection opportunities even when packet-level morphing succeeds, and we leave their joint evaluation to future work.
Although the architecture is not specific to censorship, its effectiveness has not yet been evaluated across other use cases, such as anomaly detection, website fingerprinting, or IoT traffic classification.
Our results establish the effectiveness of \sol in the evaluated censorship setting rather than universal generalization across all metadata-based MTC tasks.
Its action space is also intentionally conservative.
\sol preserves the original packet count, which may limit morphability when the target traffic has a substantially different packet structure.
Padding must also respect the effective tunnel PMTU, including encapsulation overhead, to avoid fragmentation~\cite{rfc4459}.
Added delay may overlap with natural congestion or jitter, which can complicate both deployment and detection.
Scaling to larger target sets also increases the target-selection and conditioning space and warrants further evaluation.
Our load-stress experiment further shows that implementation overhead is load dependent.
Queue occupancy increases as the post-morphing traffic demand approaches the bottleneck capacity, but the testbed remains loss-free at the evaluated near-capacity operating point.
Packet loss appears once the effective traffic demand exceeds the bottleneck capacity and increases with the degree of overload.
These measurements characterize our software-based Mininet testbed and should not be interpreted as evidence of carrier-grade hardware line-rate scalability.

Our experiments focus on TCP-based flows and early-stage windows of up to 30 packets.
Extending \sol to QUIC and other modern transports introduces additional challenges, including multiplexed streams, different loss and retransmission behavior, and application-controlled packetization.
We also evaluate history-only causal execution using a finite set of classifier decision points within the early-stage window.
Future work should evaluate \sol across multiple use cases, richer decision-time policies, larger target spaces, and defenses that combine adversarial retraining with protocol-aware features.

\section{Conclusion}
\label{sec:conclusion}

We presented \sol, a target-conditioned traffic morphing framework for evaluating the robustness of early-stage encrypted traffic classifiers. 
\sol generates concrete packet-level tunnel-layer padding and delay actions that preserve packet direction, ordering, and packet count while satisfying MTU, bandwidth, and latency constraints under a two-ended tunnel-shaping threat model.
{Our main controlled characterization uses the full-window oracle interface, which provides an upper bound on early-stage morphability rather than an online deployment setting.}
Across three datasets and four censor architectures, \sol achieves up to 85.7\% targeted evasion under this setting and outperforms prior morphing baselines under the same early-stage constraints while satisfying all specified shaping constraints.
Our analysis shows that early-stage evasion depends strongly on source--target morphability.
Positive packet-size gaps and direction compatibility make additive morphing more effective, while timing-dominated pairs leave more residual structure for adaptive censors to exploit.
{We further evaluate \sol under a history-only causal interface that supports online packet-by-packet execution without inference-time classifier queries.
Although causal morphing is substantially more challenging and environment dependent, \sol consistently outperforms the query-free BAP baseline on both VANT and UTMobileNet.}
In adaptive retraining experiments, \sol retains 43.5\% targeted evasion after four rounds.
{We further show that extending the defender's observation horizon and restricting directly manipulable features can substantially reduce evasion, revealing trade-offs between robustness, decision latency, and clean classification performance.}
{Our Mininet experiments also show that concurrent morphed traffic remains loss-free below the evaluated bottleneck capacity, while queue buildup and packet loss emerge once post-morphing traffic demand exceeds the available capacity.}
These results suggest that early-stage classifiers should be evaluated not only by clean accuracy, but also by targeted morphability under realistic shaping constraints, observation interfaces, defender adaptations, and operational network conditions.

\section*{Acknowledgment}
This research/project is supported by the National Research Foundation, Singapore, and the Cyber Security Agency of Singapore under the National Cybersecurity R\&D Programme and the CyberSG R\&D Programme Office (Award CRPO-GC2-ASTAR-001). 
Any opinions, findings, conclusions, or recommendations expressed in these materials are those of the author(s) and do not reflect the views of the National Research Foundation, Singapore, the Cyber Security Agency of Singapore, or the CyberSG R\&D Programme Office.

\bibliographystyle{ACM-Reference-Format}
\bibliography{sample-base}

\appendix

\section{Morphability Score}
\label{app:morphability_score}

For a source early-stage window $X_{\mathrm{src}}$ and a candidate target
reference $X_{\mathrm{tgt}}$, the size compatibility term penalizes
references that would require shrinking source packets or exceeding the
padding budget:
\begin{equation}
\label{eq:size_compat}
    C_s
    =
    \exp\left(
    -
    \frac{
    \sum_{i=1}^{T}[s_i^{\mathrm{src}}-s_i^{\mathrm{tgt}}]_+
    +
    \left[\sum_{i=1}^{T}[s_i^{\mathrm{tgt}}-s_i^{\mathrm{src}}]_+ - B_s\right]_+
    }{
    \sum_{i=1}^{T}s_i^{\mathrm{src}}+\epsilon
    }
    \right),
\end{equation}
where $[z]_+=\max(z,0)$, $B_s$ is the padding budget, and $\epsilon$ is a
small constant.

The timing compatibility term is defined analogously. It penalizes
references that would require packets to arrive earlier or exceed the
delay budget:
\begin{equation}
\label{eq:timing_compat}
    C_\tau = \exp\left( -
    \frac{
    \sum_{i=1}^{T}[\tau_i^{\mathrm{src}}-\tau_i^{\mathrm{tgt}}]_+ +
    \left[\sum_{i=1}^{T}[\tau_i^{\mathrm{tgt}}-\tau_i^{\mathrm{src}}]_+ - B_\tau\right]_+
    }{
    \sum_{i=1}^{T}\tau_i^{\mathrm{src}}+\epsilon
    }
    \right),
\end{equation}
where $B_\tau$ is the delay budget.

The direction compatibility term is the fraction of packet directions
that already match:
\begin{equation}
\label{eq:dir_compat}
    C_d = \frac{1}{T} \sum_{i=1}^{T}\mathbf{1}[d_i^{\mathrm{src}}=d_i^{\mathrm{tgt}}].
\end{equation}

{We further evaluate the sensitivity of the morphability score to the weighting coefficients in Eq.~\ref{eq:morphability_score}. 
Table~\ref{tab:morphability_weight_sensitivity} reports T-ESR under different allocations of $\alpha$, $\beta$, and $\gamma$ while keeping $\alpha+\beta+\gamma=1$. 
The default setting $(0.3,0.3,0.4)$ achieves the highest T-ESR of 82.4\%. 
Across all tested configurations, performance varies by at most 2.1\%, indicating that the selection mechanism is not highly sensitive to the exact weighting. 
Slightly emphasizing direction compatibility is beneficial because packet direction is immutable, whereas over-emphasizing it reduces the contribution of the controllable size and timing dimensions.}

\begin{table}[htbp]
\centering
\caption{Sensitivity of the morphability-score weights on VANT using UniNet.}
\label{tab:morphability_weight_sensitivity}
\begin{tabular}{cccc}
\toprule
$\alpha$ & $\beta$ & $\gamma$ & T-ESR (\%) \\
\midrule
0.33 & 0.33 & 0.33 & 81.6 \\
0.40 & 0.30 & 0.30 & 80.5 \\
0.30 & 0.40 & 0.30 & 80.3 \\
\textbf{0.30} & \textbf{0.30} & \textbf{0.40} & \textbf{82.4} \\
0.25 & 0.25 & 0.50 & 81.1 \\
\bottomrule
\end{tabular}
\end{table}

\section{Dataset Details and Composition}
\label{app:datasets}

\subsection{Preprocessing and Distribution}
Prior to analysis, we apply minimal data cleaning to all datasets to remove malformed traces that violate basic transport-layer semantics (e.g., incomplete handshakes or insufficient packet counts), as such flows cannot support realistic packet-level morphing.
Table~\ref{tab:app_datasets} summarizes the final class-wise distribution used in our experiments.

\begin{table}[htbp]
\centering
\caption{Detailed composition of packet-level traffic datasets used in evaluation.}
\label{tab:app_datasets}
\resizebox{0.45\columnwidth}{!}{%
\begin{tabular}{|lr|lr|}
\hline
\multicolumn{2}{|c|}{\cellcolor{gray!25}\textbf{UTMobileNet (2021)}} & \multicolumn{2}{c|}{\cellcolor{gray!25}\textbf{VANT (2024)}} \\ \hline
Google-Maps & 3184 & File Transfer & 10430 \\
Netflix     & 3021 & C2            & 9599  \\
Reddit      & 2271 & Streaming     & 1264  \\
Pinterest   & 1272 & Chat          & 1101  \\
Instagram   & 1259 & VoIP          & 617   \\
Facebook    & 1238 &                &       \\
YouTube     & 1031 & \multicolumn{2}{c|}{} \\ \cline{3-4}
Google-Drive& 956  & \multicolumn{2}{c|}{\cellcolor{gray!25}\textbf{VideoPF (2025)}} \\ \cline{3-4}
Spotify     & 971  & YouTube       & 5751  \\
Twitter     & 682  & Netflix       & 3298  \\
Gmail       & 235  & Amazon        & 4525  \\
Hangout     & 618  & Twitch        & 1066  \\
Messenger   & 394  &                &       \\ \hline
\end{tabular}%
}
\end{table}

\subsection{VideoPF Collection Methodology}
A known limitation of standard benchmarks like UTMobileNet is their reliance on controlled collection environments, which can induce overfitting to clean-room artifacts.
To complement these, our custom {VideoPF} dataset is designed to capture realistic adaptive-bitrate (ABR) video traffic under heterogeneous network conditions.
VideoPF is collected using a hybrid strategy:
\begin{itemize}
    \item \textbf{Residential Capture:} Traffic collected from residential networks to naturally include background noise and real-world ISP shaping effects.
    \item \textbf{Controlled Injection:} Traffic collected in controlled environments where we deliberately inject randomized bandwidth limits, packet loss, and latency using standard traffic control (\texttt{tc}) mechanisms.
\end{itemize}
This design prevents the classifiers (and our morpher) from overfitting to idealized network conditions, ensuring robust evaluation across diurnal variations.

\subsection{Hyperparameters}
\label{app:hparams}

\begin{table}[htbp]
\centering
\caption{\textbf{Detailed Model Architecture and Training Hyperparameters.} This table specifies the exact configuration used to generate the experimental results, aligning with the methodology defined in \S\ref{sec:architecture}.}
\label{tab:hyperparameters_optimized}
\renewcommand{\arraystretch}{1.2} 
\setlength{\tabcolsep}{8pt}
\resizebox{0.6\columnwidth}{!}{%
\begin{tabular}{ll}
\toprule
\textbf{Parameter Category} & \textbf{Configuration / Value} \\
\midrule

\multicolumn{2}{l}{\textit{\textbf{1. Traffic Representation \& Constraints}}} \\
\quad Max packets per flow ($T$) & 30 \\
\quad Input features & \{Packet size ($s_t$), Direction ($d_t$), IAT ($\tau_t$)\} \\
\quad MTU constraint (Size limit) & 1500 bytes \\
\quad Bandwidth budget & 30\% of original bytes \\
\quad Latency budget & 50\% of original flow duration \\

\midrule
\multicolumn{2}{l}{\textit{\textbf{2. Training Setup}}} \\
\quad Batch size & 128 \\
\quad Training epochs & 20 \\
\quad Optimizer & AdamW \\
\quad Learning rate & $5 \times 10^{-4}$ (with cosine decay) \\
\quad Weight decay & $1 \times 10^{-5}$ \\
\quad Computing Infrastructure & Single NVIDIA RTX 4090 (24GB VRAM) \\

\midrule
\multicolumn{2}{l}{\textit{\textbf{3. Morpher Architecture (\sol)}}} \\
\quad Core Architecture & Multi-layer Transformer Encoder + Cross-attention Decoder \\
\quad Number of layers ($L$) & 2 \\
\quad Embedding dimension ($d_{model}$) & 64 \\
\quad Feed-forward dimension ($d_{ff}$) & 128  \\
\quad Cross-attention heads ($H$) & 4 \\
\quad Dropout rate & 0.1 \\
\quad Normalization & Pre-LayerNorm \\
\quad Output activation (for $\Delta s, \Delta \tau$) & \textbf{Sigmoid} (Enforces non-negative perturbations) \\

\midrule
\multicolumn{2}{l}{\textit{\textbf{4. Target Selection Strategy}}} \\
\quad Primary criteria & Target label $\neq$ Source label \\
\quad Target class & Fixed before evaluation \\
\quad Reference pool & Evader-side training flows from the fixed target class \\
\quad {Morphability-score weights $(\alpha,\beta,\gamma)$} & {$(0.3,\,0.3,\,0.4)$} \\
\quad Selection rule & Maximize morphability score in Eq.~\ref{eq:morphability_score} \\
\quad Compatibility terms & Size, timing, and direction compatibility \\
\quad Runtime censor feedback & None \\

\midrule
\multicolumn{2}{l}{\textit{\textbf{5. Evaluation Metrics}}} \\
\quad Targeted ESR (T-ESR) & Success rate of misclassification to $y_{\text{target}}$ \\
\quad Untargeted ESR (U-ESR) & Success rate of misclassification to any $y \neq y_{\text{source}}$ \\
\quad Bandwidth overhead & $\frac{\sum \Delta s_t}{\sum s_{t, \text{orig}}} \times 100\%$ \\
\quad Latency overhead & $\frac{\text{Total Time}_{\text{morphed}} - \text{Total Time}_{\text{orig}}}{\text{Total Time}_{\text{orig}}} \times 100\%$ \\
\quad Validity rate & Percentage of flows meeting all protocol constraints \\

\bottomrule
\end{tabular}
}
\end{table}

\section{Censor Calibration Details}
\label{sec:appendix_calibration_exp1}

This appendix provides supporting details for the censor calibration summarized in \S\ref{sec:results}.

\subsection{Early-stage sequence length sensitivity.}
We analyze the impact of sequence length $T$ on detection performance. Larger values provide richer temporal context but incur linear overheads in packet buffering. We evaluate FastFlow on VideoPF while varying $T \in \{10, 20, 30, 50, 80, 100\}$.
 Detection accuracy increases rapidly up to $T{=}30$, after which gains saturate while collection latency grows sharply. At large values ($T{=}100$), performance degrades due to noise from bulk transfer phases. We select $T{=}30$ as a conservative operating point.

\subsection{Baseline classifier strength.}
Table~\ref{tab:censor_baseline} reports average classification performance across three datasets. All censors achieve Macro F1 $>$ 92\% using only the first 30 packets. UniNet achieves the highest accuracy (96.50\%) with the lowest cross-dataset variance ($\sigma{=}0.70$). The full sensitivity analysis across all four architectures is provided in Table~\ref{tab:appendix_full_results}.

\begin{table}[htbp]
\centering
\caption{Average classification performance of censors on unperturbed traffic.}
\label{tab:censor_baseline}
\begin{tabular}{l|cc}
\toprule
\textbf{Model} & Accuracy (\%) & Macro F1 (\%) \\
\midrule
Grad-BP  & 93.02 $\pm$ 1.81 & 92.48 $\pm$ 2.39 \\
GGFAST   & 95.95 $\pm$ 2.52 & 95.14 $\pm$ 3.14 \\
FastFlow & 96.35 $\pm$ 2.01 & 95.94 $\pm$ 2.16 \\
UniNet   & 96.50 $\pm$ 0.70 & 96.13 $\pm$ 0.92 \\
\bottomrule
\end{tabular}
\end{table}

\noindent\textbf{Robustness Across Censor Architectures.}
To verify that this choice is not specific to a single model, we repeat the analysis across four surrogate censors:
GGFAST, Grad-BP, UniNet, and FastFlow.
Table~\ref{tab:appendix_full_results} reports accuracy, macro F1, and inference latency as the observation window increases.

\begin{table}[t]
\centering
\caption{\textbf{Full Sensitivity Analysis.} Performance of all censor architectures under varying packet-sequence lengths.}
\label{tab:appendix_full_results}
\resizebox{0.75\linewidth}{!}{%
\begin{tabular}{c|l|cc|c}
\toprule
\textbf{Window Size ($N$)} & \textbf{Model} & \textbf{Accuracy (\%)} & \textbf{Macro F1 (\%)} & \textbf{Inference Latency (s)} \\ \midrule
\multirow{4}{*}{10} 
& Grad-BP & 88.92 & 88.25 & 0.167 \\
& GGFAST & 90.87 & 89.99 & 0.016 \\
& FastFlow & 89.34 & 88.70 & 0.013 \\
& UniNet & 83.92 & 83.50 & 0.021 \\ \midrule
\multirow{4}{*}{20} 
& Grad-BP & 94.92 & 94.91 & 0.141 \\
& GGFAST & 97.51 & 97.30 & 0.016 \\
& FastFlow & 96.98 & 96.96 & 0.014 \\
& UniNet & 94.06 & 92.75 & 0.018 \\ \midrule
\multirow{4}{*}{\textbf{30}} 
& Grad-BP & 94.54 & 94.50 & 0.212 \\
& GGFAST & {98.84} & {98.76} & {0.018} \\
& FastFlow & {98.67} & {98.43} & {0.015} \\
& UniNet & 97.19 & 97.08 & 0.021 \\ \midrule
\multirow{4}{*}{50} 
& Grad-BP & 92.49 & 91.74 & 0.327 \\
& GGFAST & 97.52 & 97.21 & 0.017 \\
& FastFlow & 98.84 & 98.56 & 0.017 \\
& UniNet & 98.77 & 98.40 & 0.018 \\ \midrule
\multirow{4}{*}{80} 
& Grad-BP & 90.51 & 90.27 & 0.452 \\
& GGFAST & 96.62 & 96.31 & 0.016 \\
& FastFlow & 97.20 & 97.15 & 0.018 \\
& UniNet & 97.38 & 97.12 & 0.018 \\ \midrule
\multirow{4}{*}{100} 
& Grad-BP & 90.07 & 89.41 & 0.601 \\
& GGFAST & 96.32 & 96.17 & 0.017 \\
& FastFlow & 98.19 & 97.87 & 0.019 \\
& UniNet & 97.72 & 97.26 & 0.019 \\ \bottomrule
\end{tabular}%
}
\end{table}

\section{Detailed Per-Pair Morphability Results}
\label{app:pair_structure_full}

\begin{table}[htbp]
\centering
\caption{Detailed structural characteristics of selected VANT morphing pairs.}
\label{tab:app_pair_structure}
\resizebox{0.8\textwidth}{!}{%
\begin{tabular}{llrrrrrrr}
\toprule
\textbf{Pair} & \textbf{Direction}
& $\bar{s}_{src}$ & $\bar{s}_{tgt}$ & $\Delta_s$
& $\bar{\tau}_{src}$ & $\bar{\tau}_{tgt}$ & $\Delta_\tau$
& \textbf{Dir-Match} \\
\midrule
A & C2$\to$Streaming
& 218 & 685 & +467 & 42.1 & 24.8 & $-17.3$ & 0.68 \\
B & C2$\to$Chat
& 218 & 234 & +16  & 42.1 & 85.4 & +43.3 & 0.38 \\
C & VoIP$\to$Chat
& 196 & 234 & +38  & 8.2  & 85.4 & +77.2 & 0.35 \\
D & VoIP$\to$Streaming
& 196 & 685 & +489 & 8.2  & 24.8 & +16.6 & 0.62 \\
\bottomrule
\end{tabular}%
}
\end{table}

\begin{table}[htbp]
\centering
\caption{Detailed static per-pair evasion performance on VANT using UniNet as the target censor. 
NG: \sol, PBB: PANTS-BB, Amo: Amoeba.}
\label{tab:app_static_perpair_full}
\resizebox{\textwidth}{!}{%
\begin{tabular}{ll|ccc|ccc|ccc|ccc}
\toprule
& &
\multicolumn{3}{c|}{\textbf{T-ESR (\%)}} &
\multicolumn{3}{c|}{\textbf{U-ESR (\%)}} &
\multicolumn{3}{c|}{\textbf{$\delta_{\mathrm{bw}}$ (\%)}} &
\multicolumn{3}{c}{\textbf{$\delta_{\mathrm{lat}}$ (\%)}} \\
\textbf{Pair} & \textbf{Direction}
& NG & PBB & Amo
& NG & PBB & Amo
& NG & PBB & Amo
& NG & PBB & Amo \\
\midrule
A & C2$\to$Streaming
& \textbf{85.8} & 65.4 & 64.2
& \textbf{91.5} & 75.2 & 78.2
& 28.5 & \textbf{28.2} & 29.4
& \textbf{30.2} & 35.8 & 48.5 \\
B & C2$\to$Chat
& \textbf{80.2} & 58.8 & 56.8
& \textbf{87.9} & 69.5 & 71.5
& \textbf{12.8} & 15.4 & 25.2
& 48.5 & \textbf{38.2} & 42.4 \\
C & VoIP$\to$Chat
& \textbf{78.5} & 56.2 & 56.5
& \textbf{89.8} & 71.6 & 73.8
& \textbf{14.5} & 18.8 & 28.5
& \textbf{45.2} & 45.5 & 48.2 \\
D & VoIP$\to$Streaming
& \textbf{85.1} & 65.5 & 62.8
& \textbf{91.6} & 77.2 & 75.7
& 27.1 & \textbf{26.5} & 29.5
& \textbf{28.4} & 32.1 & 42.8 \\
\bottomrule
\end{tabular}%
}
\end{table}

\section{Distribution of pairs}\label{app:pairdist}
\begin{figure}[htbp]
\centering
\includegraphics[width=0.8\columnwidth]{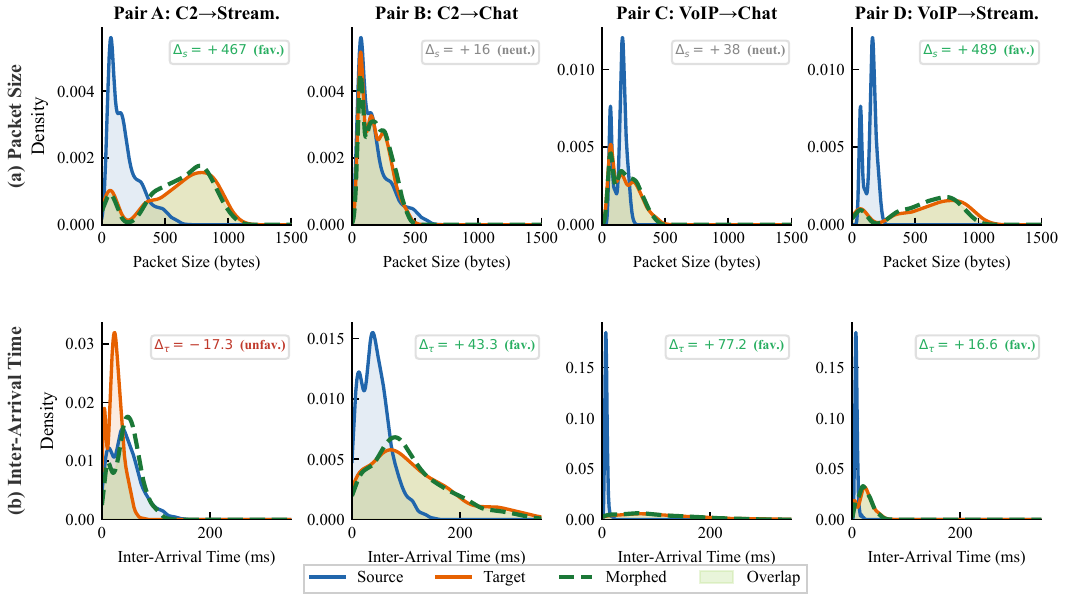}
\caption{
Per-pair distributional analysis on VANT over the first 30 packets.
\textbf{Top}: packet-size KDEs.
\textbf{Bottom}: inter-arrival-time KDEs.
\textcolor{blue}{Blue}: source.
\textcolor{orange}{Orange}: genuine target.
\textcolor{green!60!black}{Green dashed}: \sol-morphed.
}

\label{fig:pair_distribution}
\end{figure}

Fig.~\ref{fig:pair_distribution} visualizes how these overhead patterns translate into distributional changes. 
For size-dominated pairs A and D, \sol-morphed traffic closely follows the target packet-size distribution, with substantial overlap in the 550--850 byte range. 
This confirms that additive padding is effective when the target class has larger packets than the source class. 
For timing-dominated pairs B and C, source and target packet sizes are already concentrated in a similar range, so size morphing provides less discriminative leverage. 
In these cases, \sol relies more on delay to reshape the inter-arrival-time distribution.
However, delay can stretch timing patterns but cannot fully erase source-specific temporal microstructure, which explains why timing-dominated pairs remain harder than size-dominated pairs.

Overall, the pair-level analysis shows that \sol does not obtain high evasion by applying large perturbations indiscriminately. 
Instead, it adapts the type of perturbation to the source--target structure: padding is emphasized when the target has larger packets, while delay is used when packet sizes are already similar and timing carries more discriminative information. 
This motivates the morphability-aware target selection used in our history-only causal experiments: under additive-only constraints, a practical evader should prefer allowed targets that are structurally compatible with the observed source early-stage window.

\section{Details for Exp-3}\label{exp2-5-rounds}
Table~\ref{tab:multiround} reports T-ESR and U-ESR across four adaptation rounds, averaged across all VANT pairs. Both $\mathcal{G}_r$ and $\mathcal{C}_r$ are retrained at each round. $r$=0 represents the initial static setting. The T-ESR reported at round~$r$ is the performance of $\mathcal{G}_r$ evaluated against $\mathcal{C}_r$.

\begin{table}[htbp]
\centering
\caption{Multi-round adaptive evaluation on VANT. }
\label{tab:multiround}
\resizebox{0.8\columnwidth}{!}{%
\begin{tabular}{l|ccccc|ccccc}
\toprule
& \multicolumn{5}{c|}{\textbf{T-ESR (\%)}} & \multicolumn{5}{c}{\textbf{U-ESR (\%)}} \\
\textbf{Method ($G$)} & $r{=}0$ & $r{=}1$ & $r{=}2$ & $r{=}3$ & $r{=}4$ & $r{=}0$ & $r{=}1$ & $r{=}2$ & $r{=}3$ & $r{=}4$ \\
\midrule
BAP       & 18.4 & 8.2  & 5.5  & 4.8  & 4.5  & 30.5 & 18.1 & 13.2 & 11.5 & 10.8 \\
Amoeba    & 60.1 & 28.5 & 20.8 & 17.2 & 15.8 & 74.8 & 45.2 & 34.5 & 28.8 & 26.2 \\
PANTS-BB  & 61.5 & 32.8 & 23.5 & 19.4 & 17.8 & 73.4 & 48.5 & 36.2 & 30.5 & 28.1 \\
\midrule
\textbf{\sol} & \textbf{82.4} & \textbf{62.8} & \textbf{52.4} & \textbf{46.8} & \textbf{43.5} & \textbf{90.2} & \textbf{72.5} & \textbf{63.8} & \textbf{58.2} & \textbf{55.5} \\
\bottomrule
\end{tabular}%
}
\end{table}

\section{Additional Results for Exp-1}
\label{sec:appendix_exp2}

This appendix provides the comprehensive quantitative breakdown for {Exp-1}, supplementing the aggregated findings reported in \S\ref{subsec:intra_domain}. 
Specifically, we present detailed performance metrics for \sol{} and all baselines (including the state-of-the-art {PANTS-BB}~\cite{PANTS-2025-Usenix}) across two challenging environments: {UTMobileNet} (high jitter) and {VANT} (encrypted traffic).
Additionally, Table~\ref{tab:sota_benchmark} presents the precise numerical values corresponding to the visual results shown in Figure~\ref{fig:main_results_1x4} (see \S\ref{subsubsec:videopf_case_study}) for the VideoPF-2025 dataset.

\begin{table}[htbp]
\centering
\caption{\textbf{Evasion Performance on the VideoPF-2025 Dataset.}
Percentage improvements relative to the strongest baseline are highlighted in \textcolor{cyan!50!blue}{blue}. }
\label{tab:sota_benchmark}
\setlength{\tabcolsep}{5.5pt} 
\renewcommand{\arraystretch}{1.05}
\resizebox{0.8\columnwidth}{!}{%
\begin{tabular}{llccccc}
\toprule
\multirow{2}{*}{\textbf{Method}} & \multirow{2}{*}{\textbf{Target Censor}} & \multicolumn{2}{c}{\textbf{Evasion Success (ESR)}} & \textbf{Validity} & \multicolumn{2}{c}{\textbf{Overhead}} \\
\cmidrule(lr){3-4} \cmidrule(lr){6-7}
 &  & \textbf{Targeted} & \textbf{Untargeted} & \textbf{Rate} & \textbf{$\delta_{\text{bw}}$ (BW )} & \textbf{$\delta_{\text{lat}}$ (Lat)}  \\
\midrule

RandP & \multirow{5}{*}{\textbf{Grad-BP}}
 & 6.4\% & 21.5\% & 100.0\% & 24.8\% & 18.5\% \\
BAP~\cite{BAP-2021-Usenix} & 
 & 45.2\% & 62.1\% & 58.4\% & 16.5\% & 11.2\% \\
Amoeba~\cite{Amoeba-2023-CoNext} & 
 & \underline{78.5\%} & \underline{88.4\%} & 82.1\% & 28.4\% & 35.6\% \\
\textbf{PANTS-BB}~\cite{PANTS-2025-Usenix} & 
 & 68.2\% & 75.5\% & \textbf{100.0\%} & 21.8\% & 27.5\% \\
\cellcolor{gray!10}\textbf{\sol (Ours)} & 
 & \cellcolor{gray!10}\textbf{89.5\%}\tiny\textcolor{cyan!50!blue}{(+14.0\%)} & \cellcolor{gray!10}\textbf{95.2\%}\tiny\textcolor{cyan!50!blue}{(+7.7\%)} & \cellcolor{gray!10}\textbf{100.0\%} & \cellcolor{gray!10}\textbf{22.5\%} & \cellcolor{gray!10}\textbf{28.4\%} \\
\midrule

RandP & \multirow{5}{*}{\textbf{GGFAST}}
 & 4.1\% & 11.4\% & 100.0\% & 25.1\% & 20.2\% \\
BAP & 
 & 11.8\% & 24.6\% & 55.2\% & 18.2\% & 12.5\% \\
Amoeba & 
 & \underline{68.2\%} & \underline{79.5\%} & 79.5\% & 29.1\% & 42.2\% \\
\textbf{PANTS-BB} & 
 & 59.4\% & 70.1\% & \textbf{100.0\%} & 25.5\% & 31.1\% \\
\cellcolor{gray!10}\textbf{\sol (Ours)} & 
 & \cellcolor{gray!10}\textbf{84.2\%}\tiny\textcolor{cyan!50!blue}{(+23.5\%)} & \cellcolor{gray!10}\textbf{91.8\%}\tiny\textcolor{cyan!50!blue}{(+15.5\%)} & \cellcolor{gray!10}\textbf{100.0\%} & \cellcolor{gray!10}\textbf{25.2\%} & \cellcolor{gray!10}\textbf{32.1\%} \\
\midrule

RandP & \multirow{5}{*}{\textbf{FastFlow}}
 & 4.6\% & 12.8\% & 100.0\% & 24.9\% & 19.1\% \\
BAP & 
 & 32.4\% & 48.5\% & 61.3\% & 15.8\% & 12.5\% \\
Amoeba & 
 & \underline{75.6\%} & \underline{86.2\%} & 83.4\% & 27.8\% & 39.5\% \\
\textbf{PANTS-BB} & 
 & 65.5\% & 78.4\% & \textbf{100.0\%} & 29.8\% & 28.9\% \\
\cellcolor{gray!10}\textbf{\sol (Ours)} & 
 & \cellcolor{gray!10}\textbf{86.8\%}\tiny\textcolor{cyan!50!blue}{(+14.8\%)} & \cellcolor{gray!10}\textbf{93.5\%}\tiny\textcolor{cyan!50!blue}{(+8.5\%)} & \cellcolor{gray!10}\textbf{100.0\%} & \cellcolor{gray!10}\textbf{23.8\%} & \cellcolor{gray!10}\textbf{36.5\%} \\
\midrule

RandP & \multirow{5}{*}{\textbf{UniNet}}
 & 2.8\% & 12.4\% & 100.0\% & 25.0\% & 19.8\% \\
BAP & 
 & 25.1\% & 40.2\% & 59.8\% & 17.5\% & 13.6\% \\
Amoeba & 
 & {72.4\%} & {84.1\%} & 81.2\% & 29.4\% & 45.9\% \\
\textbf{PANTS-BB} & 
 & \underline{74.1}\% & \underline{86.7}\% & \textbf{100.0\%} & 26.5\% & 23.2\% \\
\cellcolor{gray!10}\textbf{\sol (Ours)} & 
 & \cellcolor{gray!10}\textbf{82.4\%}\tiny\textcolor{cyan!50!blue}{(+11.2\%)} & \cellcolor{gray!10}\textbf{90.6\%}\tiny\textcolor{cyan!50!blue}{(+4.5\%)} & \cellcolor{gray!10}\textbf{100.0\%} & \cellcolor{gray!10}\textbf{26.1\%} & \cellcolor{gray!10}\textbf{40.5\%} \\

\bottomrule
\end{tabular}%
}
\end{table}

\begin{table*}[htbp]
\centering
\footnotesize

\begin{minipage}[t]{0.49\textwidth}
\centering
\caption{\textbf{Detailed Evasion Results on UTMobileNet.}
Breakdown by target censor.}
\label{tab:utmobile_results}
\setlength{\tabcolsep}{2.5pt}
\renewcommand{\arraystretch}{1.05}
\resizebox{\linewidth}{!}{%
\begin{tabular}{llccccc}
\toprule
\multirow{2}{*}{\textbf{Censor}} & \multirow{2}{*}{\textbf{Method}} &
\multicolumn{2}{c}{\textbf{ESR (\%)}} &
\textbf{Valid.} &
\multicolumn{2}{c}{\textbf{Overhead (\%)}} \\
\cmidrule(lr){3-4} \cmidrule(lr){6-7}
 &  & \textbf{T} & \textbf{U} & \textbf{Rate} & \textbf{BW} & \textbf{Lat} \\
\midrule

\multirow{5}{*}{\textbf{Grad-BP}} 
 & RandP            & 3.5  & 12.4 & 100.0 & 28.5 & 22.1 \\
 & BAP              & 21.2 & 34.6 & 52.1  & 20.5 & 14.8 \\
 & Amoeba           & 61.8 & 75.2 & 79.5  & 29.2 & 38.4 \\
 & \textbf{PANTS-BB} & 55.4 & 68.5 & \textbf{100.0} & \textbf{21.2} & 30.5 \\
 & \cellcolor{gray!10}\textbf{\sol}
   & \cellcolor{gray!10}\textbf{82.5}
   & \cellcolor{gray!10}\textbf{90.4}
   & \cellcolor{gray!10}\textbf{100.0}
   & \cellcolor{gray!10}25.8
   & \cellcolor{gray!10}\textbf{30.2} \\
\midrule

\multirow{5}{*}{\textbf{GGFAST}} 
 & RandP            & 2.8  & 9.5  & 100.0 & 29.1 & 23.5 \\
 & BAP              & 15.5 & 28.2 & 48.6  & 21.8 & 16.2 \\
 & Amoeba           & 58.4 & 71.1 & 76.2  & 29.5 & 44.5 \\
 & \textbf{PANTS-BB} & 49.2 & 64.1 & \textbf{100.0} & \textbf{22.5} & 32.1 \\
 & \cellcolor{gray!10}\textbf{\sol}
   & \cellcolor{gray!10}\textbf{79.5}
   & \cellcolor{gray!10}\textbf{87.8}
   & \cellcolor{gray!10}\textbf{100.0}
   & \cellcolor{gray!10}27.2
   & \cellcolor{gray!10}\textbf{33.8} \\
\midrule

\multirow{5}{*}{\textbf{FastFlow}} 
 & RandP            & 3.1  & 11.6 & 100.0 & 28.8 & 22.8 \\
 & BAP              & 29.5 & 44.1 & 55.4  & 19.2 & 15.1 \\
 & Amoeba           & 65.8 & 78.5 & 81.8  & 28.4 & 41.2 \\
 & \textbf{PANTS-BB} & 53.6 & 70.4 & \textbf{100.0} & \textbf{21.8} & 31.4 \\
 & \cellcolor{gray!10}\textbf{\sol}
   & \cellcolor{gray!10}\textbf{83.6}
   & \cellcolor{gray!10}\textbf{91.2}
   & \cellcolor{gray!10}\textbf{100.0}
   & \cellcolor{gray!10}26.5
   & \cellcolor{gray!10}\textbf{38.1} \\
\midrule

\multirow{5}{*}{\textbf{UniNet}} 
 & RandP            & 3.4  & 16.8 & 100.0 & 29.5 & 24.2 \\
 & BAP              & 22.4 & 36.5 & 51.2  & 22.5 & 17.6 \\
 & Amoeba           & 70.1 & 81.4 & 78.5  & 27.2 & 48.6 \\
 & \textbf{PANTS-BB} & \underline{74.5} & \underline{82.8} & \textbf{100.0} & \textbf{24.2} & 35.5 \\
 & \cellcolor{gray!10}\textbf{\sol}
   & \cellcolor{gray!10}\textbf{81.5}
   & \cellcolor{gray!10}\textbf{89.2}
   & \cellcolor{gray!10}\textbf{100.0}
   & \cellcolor{gray!10}29.4
   & \cellcolor{gray!10}\textbf{42.5} \\
\bottomrule
\end{tabular}%
}
\end{minipage}
\hfill
\begin{minipage}[t]{0.49\textwidth}
\centering
\caption{\textbf{Detailed Evasion Results on VANT.}
Breakdown by target censor.}
\label{tab:vant_results}
\setlength{\tabcolsep}{2.5pt}
\renewcommand{\arraystretch}{1.05}
\resizebox{\linewidth}{!}{%
\begin{tabular}{llccccc}
\toprule
\multirow{2}{*}{\textbf{Censor}} & \multirow{2}{*}{\textbf{Method}} &
\multicolumn{2}{c}{\textbf{ESR (\%)}} &
\textbf{Valid.} &
\multicolumn{2}{c}{\textbf{Overhead (\%)}} \\
\cmidrule(lr){3-4} \cmidrule(lr){6-7}
 &  & \textbf{T} & \textbf{U} & \textbf{Rate} & \textbf{BW} & \textbf{Lat} \\
\midrule

\multirow{5}{*}{\textbf{Grad-BP}} 
 & RandP            & 4.5  & 11.2 & 100.0 & 25.2 & 28.5 \\ 
 & BAP              & 19.6 & 31.4 & 48.5  & 25.6 & 18.2 \\ 
 & Amoeba           & 60.4 & 73.5 & 75.4  & 28.5 & 45.6 \\
 & \textbf{PANTS-BB} & 58.2 & 71.1 & \textbf{100.0} & \textbf{26.5} & 35.8 \\
 & \cellcolor{gray!10}\textbf{\sol}
   & \cellcolor{gray!10}\textbf{83.5}
   & \cellcolor{gray!10}\textbf{91.2}
   & \cellcolor{gray!10}\textbf{100.0}
   & \cellcolor{gray!10}27.1
   & \cellcolor{gray!10}\textbf{38.5} \\ 
\midrule

\multirow{5}{*}{\textbf{GGFAST}} 
 & RandP            & 3.5  & 9.8  & 100.0 & 26.1 & 30.2 \\
 & BAP              & 15.5 & 28.2 & 45.2  & 26.8 & 20.5 \\
 & Amoeba           & 58.5 & 70.1 & 72.1  & 26.4 & 48.1 \\
 & \textbf{PANTS-BB} & 55.5 & 68.4 & \textbf{100.0} & \textbf{27.8} & 38.5 \\
 & \cellcolor{gray!10}\textbf{\sol}
   & \cellcolor{gray!10}\textbf{81.8}
   & \cellcolor{gray!10}\textbf{89.6}
   & \cellcolor{gray!10}\textbf{100.0}
   & \cellcolor{gray!10}28.5
   & \cellcolor{gray!10}\textbf{44.2} \\
\midrule

\multirow{5}{*}{\textbf{FastFlow}} 
 & RandP            & 4.2  & 12.5 & 100.0 & 25.8 & 29.4 \\
 & BAP              & 20.4 & 35.1 & 51.5  & 24.2 & 19.1 \\
 & Amoeba           & 62.5 & 76.2 & 78.6  & 29.1 & 48.5 \\
 & \textbf{PANTS-BB} & 61.8 & 74.5 & \textbf{100.0} & \textbf{26.2} & 39.1 \\
 & \cellcolor{gray!10}\textbf{\sol}
   & \cellcolor{gray!10}\textbf{83.2}
   & \cellcolor{gray!10}\textbf{91.4}
   & \cellcolor{gray!10}\textbf{100.0}
   & \cellcolor{gray!10}22.8
   & \cellcolor{gray!10}\textbf{46.8} \\
\midrule

\multirow{5}{*}{\textbf{UniNet}} 
 & RandP            & 3.1  & 10.5 & 100.0 & 26.5 & 31.5 \\
 & BAP              & 18.5 & 32.4 & 44.6  & 28.5 & 22.1 \\
 & Amoeba           & 65.4 & 77.8 & 74.2  & 25.6 & 48.5 \\
 & \textbf{PANTS-BB} & \underline{70.8} & \underline{79.5} & \textbf{100.0} & \textbf{28.5} & 41.2 \\
 & \cellcolor{gray!10}\textbf{\sol}
   & \cellcolor{gray!10}\textbf{79.5}
   & \cellcolor{gray!10}\textbf{88.2}
   & \cellcolor{gray!10}\textbf{100.0}
   & \cellcolor{gray!10}28.6
   & \cellcolor{gray!10}\textbf{46.2} \\
\bottomrule
\end{tabular}%
}
\end{minipage}

\end{table*}

\section{Additional Results for History-Only Causal Evaluation (Exp-2)}
\label{app:utmobile-online}

\begin{table}[htbp]
    \centering
    \caption{{History-only causal attack effectiveness on UTMobileNet-2021.}}
    \label{tab:utmobile-effectiveness}
    \scriptsize
    \setlength{\tabcolsep}{3pt}
    \renewcommand{\arraystretch}{0.95}
    \resizebox{0.6\columnwidth}{!}{
    \begin{tabular}{c ccc ccc}
        \toprule
        \multirow{2}{*}{\textbf{Decision} $k$}
        & \multicolumn{3}{c}{\textbf{T-ESR (\%)}}
        & \multicolumn{3}{c}{\textbf{U-ESR (\%)}} \\
        \cmidrule(lr){2-4}\cmidrule(lr){5-7}
        & {BAP} & Amoeba & \sol
        & {BAP} & Amoeba & \sol \\
        \midrule
         5 & 1.84 & 2.18 & \textbf{4.35}
           & 24.80 & 55.15 & \textbf{56.17} \\
        10 & 2.26 & 1.16 & \textbf{5.81}
           & 29.53 & 37.16 & \textbf{43.11} \\
        15 & 3.37 & 1.31 & \textbf{5.22}
           & 30.18 & 24.82 & \textbf{38.17} \\
        20 & \textbf{5.81} & 0.73 & 5.52
           & 32.02 & 22.79 & \textbf{33.24} \\
        30 & \textbf{7.28} & 2.76 & 7.11
           & 34.57 & 31.79 & \textbf{38.75} \\
        \midrule
        Mean & 4.11 & 1.63 & \textbf{5.60}
             & 30.22 & 34.34 & \textbf{41.89} \\
        \bottomrule
    \end{tabular}}
\end{table}

\begin{table}[htbp]
    \centering
    \caption{{History-only causal efficiency comparison on UTMobileNet-2021.}}
    \label{tab:utmobile-efficiency}
    \scriptsize
    \setlength{\tabcolsep}{3pt}
    \renewcommand{\arraystretch}{0.95}
    \resizebox{0.68\columnwidth}{!}{
    \begin{tabular}{l ccc}
        \toprule
        \textbf{Metric}
        & \textbf{{BAP}}
        & \textbf{Amoeba}
        & \textbf{\sol} \\
        \midrule
        Bandwidth overhead (\%)
            & 15.8 & 24.52 & \textbf{13.53} \\
        Latency overhead (\%)
            & \textbf{21.3} & 37.17 & 44.35 \\
        Generation latency (ms/flow)
            & 35.5 & 114.98 & \textbf{0.54} \\
        Classifier queries at runtime
            & None & Per packet & None \\
        \bottomrule
    \end{tabular}}
\end{table}

\section{Ablation Studies}
\label{sec:ablation}

We provide additional ablation and robustness analysis to isolate the main design choices and evaluate the supplementary settings requested in our robustness study. The target-reference selection ablation is evaluated under a static UniNet censor, while the conditioning ablation reports both static ($C_0$) and single-round adaptive ($C_1$) performance.

\subsection{Ablation 1: Morphability-Aware Target-Reference Selection}
\label{sec:ablation_target_selection}

We evaluate the effect of morphability-aware target-reference selection
(\S\ref{sec:target-selection}). 
For each evaluation trial, the target class is fixed before morphing and is shared across all variants. 
The selection mechanism only chooses a target reference within this fixed target class. 
It uses evader-side training data and source-window statistics, including packet-size compatibility, timing compatibility, and direction-sequence compatibility. 
It does not use test-time attack success, deployed-censor outputs, confidence scores, gradients, or runtime queries.
To isolate its effect, we replace morphability-aware reference selection with
\textbf{Random} reference selection. 
For each source flow and fixed target class, the target reference is sampled uniformly from the evader-side reference pool of that target class. 
All other components of \sol remain unchanged. 
We report per-pair results on VANT under a static UniNet censor. 
This ablation does not change the targeted-evasion objective; instead, it quantifies how much targeted early-stage morphing benefits from selecting a structurally compatible reference under additive-only shaping constraints.

\begin{table*}[htbp]
\centering

\begin{minipage}[t]{0.46\textwidth}
\centering
\caption{Ablation 1: Effect of morphability-aware target-reference selection on VANT using UniNet.}
\label{tab:ablation_target_selection}
\setlength{\tabcolsep}{4.2pt}
\renewcommand{\arraystretch}{1.08}
\resizebox{\linewidth}{!}{%
\begin{tabular}{cl|cc|cc}
\toprule
& & \multicolumn{2}{c|}{\textbf{T-ESR}} &
\multicolumn{2}{c}{\textbf{U-ESR}} \\
\textbf{Pair} & \textbf{Direction}
& Random Ref. & MA Ref.
& Random Ref. & MA Ref. \\
\midrule
A & C2$\to$Streaming   & 78.2 & \textbf{85.8} & 86.5 & \textbf{91.5} \\
B & C2$\to$Chat        & 62.4 & \textbf{80.2} & 74.5 & \textbf{87.9} \\
C & VoIP$\to$Chat      & 58.8 & \textbf{78.5} & 72.1 & \textbf{89.8} \\
D & VoIP$\to$Streaming & 76.5 & \textbf{85.1} & 85.8 & \textbf{91.6} \\
\midrule
\multicolumn{2}{c|}{Average}
& 68.9 & \textbf{82.4}
& 79.7 & \textbf{90.2} \\
\bottomrule
\end{tabular}%
}
\end{minipage}
\hfill
\begin{minipage}[t]{0.52\textwidth}
\centering
\caption{Ablation 2: Effect of the cross-attention mechanism on VANT using UniNet.}
\label{tab:ablation_ca}
\setlength{\tabcolsep}{3.5pt}
\renewcommand{\arraystretch}{1.08}
\resizebox{\linewidth}{!}{%
\begin{tabular}{l|cc|cc|cc|c}
\toprule
& \multicolumn{2}{c|}{\textbf{T-ESR}} 
& \multicolumn{2}{c|}{\textbf{U-ESR}} 
& \multicolumn{2}{c|}{\textbf{Gap}} 
& \textbf{Valid.} \\
\textbf{Variant} 
& $\mathcal{C}_0$ & $\mathcal{C}_1$ 
& $\mathcal{C}_0$ & $\mathcal{C}_1$ 
& $\mathcal{C}_0$ & $\mathcal{C}_1$ 
& \scriptsize{(\%)} \\
\midrule
Target-Label  
& 58.2 & 34.5 
& 79.8 & 55.2 
& 21.6 & 20.7 
& 100 \\
MLP-Concat    
& 66.5 & 42.8 
& 83.4 & 60.5 
& 16.9 & 17.7 
& 100 \\
\textbf{Cross-Attn.} 
& \textbf{79.5} & \textbf{58.4} 
& \textbf{88.2} & \textbf{68.8} 
& \textbf{8.7} & \textbf{10.4} 
& \textbf{100} \\
\bottomrule
\end{tabular}%
}
\end{minipage}

\end{table*}

Table~\ref{tab:ablation_target_selection} shows that morphability-aware reference selection improves average T-ESR from 68.9\% to 82.4\%, and
average U-ESR from 79.7\% to 90.2\%. 
These correspond to absolute gains of 13.5 and 10.5\%, respectively. 
The improvement confirms that, even when the target class is fixed, the choice of target reference matters under additive-only shaping. 
This is expected: the morpher can add padding and delay, but cannot shrink packets, advance packets in time, change directions, reorder packets, or modify packet count. 
Some references within the same target class are therefore structurally easier to realize than others.
The gains are pair-dependent. 
For size-dominated and direction-compatible pairs such as C2$\to$Streaming and VoIP$\to$Streaming, random reference selection already finds compatible references relatively often, so the T-ESR gains are moderate: 7.6\% and 8.6\%, respectively. 
These pairs have large positive packet-size gaps and higher direction compatibility, allowing padding to move source traffic toward the target distribution.
The effect is larger for timing-dominated or direction-mismatched pairs, such as C2$\to$Chat and VoIP$\to$Chat. 
Their T-ESR improves by 17.8 and 19.7\%, respectively. 
In these cases, randomly selected references are more likely to introduce structural mismatches that the morpher cannot repair, especially mismatched packet directions and timing patterns. 
Morphability-aware reference selection avoids such references and allows the morpher to spend its limited bandwidth and latency budgets on controllable dimensions: packet size and inter-arrival time.

\begin{figure}[htbp]
    \centering
    \includegraphics[width=0.5\linewidth]{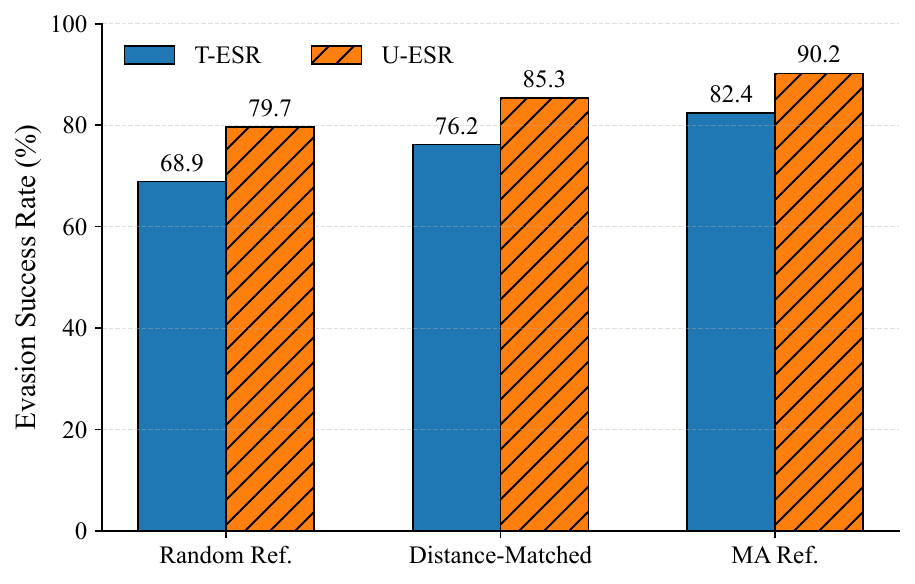}
    \caption{Target-reference selection on VANT using UniNet.}
    
    \label{fig:distance_matched_control}
\end{figure}

{\textbf{Distance-matched control.} 
To determine whether the gain from morphability-aware reference selection simply results from choosing a target reference that is closer to the source, we construct a distance-matched control.
For a source window $X_{\mathrm{src}}$ and candidate reference $X_{\mathrm{ref}}$, we define a symmetric source--reference distance as the $L_1$ distance over the normalized packet-size, inter-arrival-time, and direction features.
For each source flow and fixed target class, we first obtain the reference selected by the morphability-aware (MA) selector and then choose an alternative reference whose $L_1$ distance to the source is as close as possible to that of the MA-selected reference, without using the asymmetric feasibility score in Eq.~\ref{eq:morphability_score}.
This control therefore preserves overall source--reference proximity while removing the explicit preference for references that are feasible under additive padding, non-negative delay, and immutable packet directions.
}

{As shown in Fig.~\ref{fig:distance_matched_control}, random reference selection achieves 68.9\% T-ESR, while the distance-matched control improves it to 76.2\%, indicating that source--reference proximity itself contributes to successful morphing.
However, MA selection further increases T-ESR to 82.4\%, yielding an additional 6.2\% gain over the distance-matched control.
The same trend is observed for U-ESR, which increases from 79.7\% to 85.3\% and further to 90.2\%.
These results indicate that beyond controlling for symmetric source--reference distance, explicitly modeling the asymmetric feasibility of the permitted shaping operations provides additional benefit.}

These results support the motivation for morphability-aware target-reference selection. 
They also clarify its interpretation: the target class remains fixed, and the ablation measures the benefit of selecting a feasible reference within that pre-specified class. 
It is therefore not a comparison between easy and hard target classes, but a comparison between random and morphability-aware references under the same targeted-evasion objective.

\subsection{Ablation 2: Cross-Attention Mechanism}
\label{sec:ablation_ca}

We replace the cross-attention module (Eq.~\ref{eq:attention}) with two simpler conditioning strategies: (i)~\textbf{MLP-Concat}, which concatenates the mean-pooled target embedding with each source timestep and processes the result through a two-layer MLP; and (ii)~\textbf{Target-Label}, which conditions on a learned class embedding (a single vector per target class) rather than a full target flow.
Both variants retain the same Sigmoid output layer, training objective, and morphability-aware target-reference selection strategy.
Table~\ref{tab:ablation_ca} compares the performance of the three conditioning mechanisms on the VANT dataset.

\noindent\textbf{Analysis.} The cross-attention mechanism contributes the largest single improvement, adding 13.0\% of T-ESR over MLP-Concat and 21.3\% over Target-Label under the static censor. The most revealing metric is the \emph{gap between U-ESR and T-ESR}, which measures the difficulty of steering morphed traffic toward a specific target class versus merely escaping the source class. Target-Label conditioning produces a gap of $\sim$21\%: the generator learns a coarse per-class perturbation pattern but cannot adapt to the structural variation within the target class. MLP-Concat reduces this gap to $\sim$17\% by incorporating flow-level target information, but mean-pooling discards the temporal structure of the target flow. Cross-attention reduces the gap to $<$11\% by allowing the generator to \emph{query} the target flow at each timestep, dynamically aligning source and target at the packet level.

Under adaptation ($\mathcal{C}_1$), the advantage of cross-attention grows: the T-ESR gap over MLP-Concat widens from 13.0 to 15.6\%. This is because cross-attention produces a richer, more diverse set of per-flow transformations (each source flow queries a different target reference), making the morphing distribution harder for the adaptive classifier to cover compactly. In contrast, MLP-Concat and Target-Label produce more uniform perturbation patterns that the classifier can learn to recognize after a single retraining round.

\subsection{{Ablation 3: White-box Upper-Bound Evaluation}}\label{app:whitebox_upper}

To quantify the effect of surrogate mismatch, we additionally evaluate each trained morpher against the exact surrogate classifier used during training. 
Unlike the cross-classifier results in Section~\ref{subsec:transferability}, this setting removes cross-classifier mismatch and provides a white-box upper bound on early-stage morphability. 
As shown in Table~\ref{tab:whitebox_upper}, the transfer gap ranges from 5.8 to 14.4\%, indicating that surrogate mismatch reduces attack effectiveness.
The consistently high white-box T-ESR (89.2--93.9\%) demonstrates substantial intrinsic early-stage morphability.

\begin{table}[htbp]
\centering
\caption{{White-box upper bound and transfer gap in T-ESR (\%).}}
\label{tab:whitebox_upper}
\resizebox{0.7\columnwidth}{!}{
\begin{tabular}{lccc|ccc}
\toprule
& \multicolumn{3}{c|}{FastFlow} & \multicolumn{3}{c}{UniNet} \\
Dataset & WB & Transfer & Gap (\%) & WB & Transfer & Gap (\%) \\
\midrule
VideoPF     & 89.2 & 81.8 & 7.4 & 93.4 & 82.4 & 11.0 \\
UTMobileNet & 91.4 & 85.6 & 5.8 & 92.7 & 81.5 & 11.2 \\
VANT        & 90.6 & 83.2 & 7.4 & 93.9 & 79.5 & 14.4 \\
\bottomrule
\end{tabular}}
\end{table}

\subsection{{Budget-Allocation Analysis}}
\label{app:budget_allocation}

The default projection in Eqs.~\ref{eq:size_projection}--\ref{eq:delay_projection} applies actions sequentially and updates the remaining bandwidth and latency budgets after each packet. 
This design is simple and directly guarantees the cumulative budget constraints, but it may allow earlier packets to consume a larger fraction of the available budget.
To quantify whether this sequential allocation limits attack effectiveness, we additionally consider a full-window hindsight projection. 
Let $w_t^s=\sigma(o_t^s)$ and $w_t^\tau=\sigma(o_t^\tau)$ denote the decoder-derived padding and delay weights. Using all $T$ packet scores jointly, the delay budget is allocated as
\begin{equation}
\Delta \tau_t^{\mathrm{H}} = B_\tau \frac{w_t^\tau} {\sum_{j=1}^{T} w_j^\tau}.
\label{eq:hindsight_delay}
\end{equation}
For padding, the allocation must additionally respect the packet-specific
MTU headroom:
\begin{equation}
\Delta s_t^{\mathrm{H}} = \min\left\{ M-s_t,\; \lambda_s w_t^s \right\},
\label{eq:hindsight_padding}
\end{equation}
where $\lambda_s$ is chosen such that
$\sum_{t=1}^{T}\Delta s_t^{\mathrm{H}}\leq B_s$.
If a packet reaches its MTU limit, the unused budget is redistributed among the remaining packets. 
Thus, unlike the sequential projection, the hindsight variant allocates the available budgets jointly over the complete early-stage window while preserving the same total bandwidth, latency, and MTU constraints.

\begin{table}[htbp]
    \centering
    \caption{Effect of budget-allocation strategy on VANT using UniNet.}
    \label{tab:budget_allocation}
    \resizebox{0.5\linewidth}{!}{%
    \begin{tabular}{lcccc}
        \toprule
        \textbf{Projection} &
        \textbf{T-ESR} &
        \textbf{U-ESR} &
        $\boldsymbol{\delta_{\mathrm{bw}}}$ &
        $\boldsymbol{\delta_{\mathrm{lat}}}$ \\
        & (\%) & (\%) & (\%) & (\%) \\
        \midrule
        Sequential (Eqs.~6--7) & 82.4 & 90.2 & \textbf{27.6} & \textbf{39.4} \\
        Global Hindsight       & \textbf{83.6} & \textbf{92.1} & 28.4 & 39.8 \\
        \bottomrule
    \end{tabular}%
    }
\end{table}

As shown in Table~\ref{tab:budget_allocation}, the hindsight allocation increases T-ESR from 82.4\% to 83.6\% and U-ESR from 90.2\% to 92.1\%. 
The improvement is accompanied by slightly higher bandwidth and latency overheads, increasing from 27.6\% to 28.4\% and from 39.4\% to 39.8\%, respectively. 
These results show that globally reallocating the available budget with full-window hindsight provides only a modest improvement. 
Thus, although sequential projection can favor earlier packets, it does not substantially limit \sol's full-window morphability under the evaluated budgets.

\subsection{{Defenses Beyond the Default Early-Stage Classifier}}
\label{sec:defense_features}

We further examine two defenses that change the censor itself rather than retraining it on morphed traffic. 
The first extends the observation window beyond the 30 packets modified by \sol. 
The second removes features that can be directly affected by padding or delay.

\subsubsection{{Longer Observation Windows}}

We first allow the defender to observe beyond the 30-packet attack window, while \sol continues to modify only the first 30 packets.
Packets after position 30 are left unchanged. Table~\ref{tab:longer-window} reports the resulting clean classification performance and evasion rates.

\begin{table}[htbp]
    \centering
    \caption{Effect of extending the defender's observation window on VANT. \sol modifies only the first 30 packets; later packets remain unchanged.}
    \label{tab:longer-window}

    \resizebox{0.65\linewidth}{!}{%
    \begin{tabular}{c ccc}
        \toprule
        \textbf{Defender's observation window} &
        \textbf{Clean F1 (\%)} &
        \textbf{T-ESR (\%)} &
        \textbf{U-ESR (\%)} \\
        \midrule
        30  & 97.2 & 82.4 & 90.2 \\
        50  & \textbf{98.8} & 43.6 & 68.7 \\
        100 & 97.7 & \textbf{26.1} & \textbf{46.4} \\
        \bottomrule
    \end{tabular}%
    }
\end{table}

\subsubsection{{Feature-Restricted Defenders}}

We next evaluate classifiers that remove features directly affected by \sol's shaping actions. We consider the original representation
(packet size, direction, and IAT), a representation without packet size (direction and IAT), and a direction-only representation.

\begin{table}[htbp]
    \centering
    \caption{Effect of restricting the defender's input features on VANT. Lower ESR indicates greater robustness to \sol.}
    \label{tab:feature-defense}

    \resizebox{0.65\linewidth}{!}{%
    \begin{tabular}{l ccc}
        \toprule
        \textbf{Defender's input features} &
        \textbf{Clean F1 (\%)} &
        \textbf{T-ESR (\%)} &
        \textbf{U-ESR (\%)} \\
        \midrule
        Size + Dir. + IAT & \textbf{97.2} & 82.4 & 90.2 \\
        Dir. + IAT        & 62.1 & 42.3 & 56.4 \\
        Direction only    & 41.5 & \textbf{2.74} & \textbf{7.41} \\
        \bottomrule
    \end{tabular}%
    }
\end{table}

\end{document}